\documentclass[12pt]{article}
\usepackage{amssymb, epsfig,amssymb, latexsym}
\usepackage{amsfonts,psfrag,amsmath,bbm,color,multirow}
\usepackage{placeins} 

\graphicspath{{./report-xx-092315/}}

\newtheorem{remark}{Remark}[section]

\def\PP{{{\rm l}\kern - .15em {\rm P} }}
\def\PN2{{\PP_{N}-\PP_{N-2}}}

\newcommand{\R}{\mathbbm{R}}

\newcommand{\bM}{{\boldsymbol M}}

\newcommand{\bS}{{\boldsymbol S}}
\newcommand{\bu}{{\boldsymbol u}}
\newcommand{\bU}{{\boldsymbol U}}

\newcommand{\bw}{{\boldsymbol w}}
\newcommand{\bx}{{\bf x}}
\newcommand{\bX}{{\bf X}}
\newcommand{\bfeta}{\boldsymbol{\eta}}
\newcommand{\bphi}{\boldsymbol{\varphi}}
\newcommand{\btau}{\boldsymbol{\tau}}

\newcommand{\op}{\overline{p}}
\newcommand{\ou}{\overline{u}}
\newcommand{\oU}{\overline{U}}
\newcommand{\obu}{\overline{\boldsymbol u}}
\newcommand{\obU}{\overline{\boldsymbol U}}

\newcommand{\mK}{\mathcal{K}}
\newcommand{\mO}{\mathcal{O}}

\newcommand{\deleted}[1]{{}}

\begin{document}

\title{Approximate Deconvolution \\ Reduced Order Modeling}

\author{Xuping Xie, David Wells, Zhu Wang, Traian Iliescu}


\date{\today}

\maketitle

\begin{abstract}
This paper proposes a large eddy simulation reduced order model (LES-ROM) framework for the numerical simulation of realistic flows.
In this LES-ROM framework, the proper orthogonal decomposition (POD) is used to define the ROM basis and a POD differential filter is used to define the large ROM structures. 
An approximate deconvolution (AD) approach is used to solve the ROM closure problem and develop a new AD-ROM.
This AD-ROM is tested in the numerical simulation of the one-dimensional Burgers equation with a small diffusion coefficient ($\nu=10^{-3}$). 
\end{abstract}

\medskip
{\bf Key Words:} Reduced order models, proper orthogonal decomposition, approximate deconvolution.

\medskip
{\bf Mathematics Subject Classifications (2000)}: 65M15, 65M60

\section{Introduction}
\label{sec:introduction}

The direct numerical simulation (DNS) of realistic turbulent flow is generally extremely challenging.
The main hurdle is the enormous range of spatial and temporal scales that need to be approximated.
Indeed, capturing all the energy containing scales, down to the Kolomogorov scale, requires a number of degrees of freedom that scales like $Re^{9/4}$, where $Re$ is the Reynolds number~\cite{Sag06}.
Thus, alternative numerical methods are generally used in practice.

One of the most successful approaches to the numerical simulation of realistic turbulent flows is large eddy simulation (LES)~\cite{BIL05,Sag06}.
The idea in LES is straightforward.
Instead of approximating the flow variables, the LES models aim at approximating the spatially filtered flow variables, which can be discretized on significantly coarser spatial meshes. 
However, since the underlying Navier Stokes equations (NSE) are nonlinear, filtering them yields a system of equations that is not closed.
Closing the spatially filtered NSE, which is commonly known as the {\it closure problem}, represents the main challenge faced by LES.

There are many different closure models used in LES.
Following Sagaut's terminology, these LES closure models could be divided into two categories:
The {\it functional closure models} (chapters 5 and 6 in~\cite{Sag06}) follow from phenomenological arguments and aim at modeling the physical effect of the sub-filter scales (i.e., the scales below the below the spatial filter radius).
The main tool used in developing these functional closure models is Kolmogorov's statistical theory of turbulence and the resulting energy cascade, which postulates that the main role of the sub-filter scales is to drain energy out of the LES model.
In the functional closure models, this is generally achieved by adding an eddy viscosity to the molecular viscosity of the system.
Probably the most popular example in this class is the Smagorinsky model~\cite{Sma63}.
The second type of LES closure models is {\it structural closure models} (chapter 7 in~\cite{Sag06}).
These closure models are generally derived through mathematical rather than phenomenological arguments, e.g., formal series expansions.
One of the most popular models in this class is the approximate deconvolution (AD) model~\cite{SA99,SAK01a,SAK01b}, which uses the deconvolution approaches developed in the image processing and inverse problems communities to recover the original signal from a blurred filtered signal.

Another DNS alternative to the numerical simulation of fluid flows is centered around reduced order models (ROMs), which can reduce the computational time of a DNS by orders of magnitude without significantly reducing its accuracy.
The proper orthogonal decomposition (POD) is one of the most successful approaches for ROM development. 
An accurate numerical simulation is used in POD to extract the dominant structures, which are then used in a Galerkin approximation of the underlying equations~\cite{HLB96,Sir87abc}. 
In this paper, POD will be exclusively used to construct the ROMs.

Standard ROMs are extremely efficient and relatively accurate for laminar flows.
They generally fail, however, in the numerical simulation of realistic turbulent flows~\cite{AHLS88,balajewicz2013low,ballarin2015supremizer,giere2015supg,pacciarini2014stabilized,quarteroni2011certified,wang2012proper}.  
The main challenge that ROMs face in these realistic settings is the same as that faced by LES models: the closure problem.
Indeed, to ensure a low computational cost, only the first few POD modes are generally used in the ROM.
The resulting low-dimensional ROM, however, generally yields poor results in the numerical simulation of convection-dominated flows, often in the form of numerical oscillations.
For example, in~\cite{wang2012proper} it was shown that, for a three-dimensional (3D) flow past a cylinder at $Re=10^3$, the standard ROM results were inaccurate: they over-predicted the coherent structures, kinetic energy content, and POD mode coefficient evolution. 
The results in~\cite{wang2012proper} (see also~\cite{lassila2014model,noack2011reduced}) clearly show that the effect of the POD modes that are not used in the ROM needs to be modeled, i.e., {\it the ROM closure problem needs to be solved}.
Over the years, numerous ROM closure models have been devised (see, e.g.,~\cite{amsallem2012stabilization,AHLS88,balajewicz2013low,ballarin2015supremizer,barone2009stable,bergmann2009enablers,quarteroni2011certified,wang2012proper}).
Although a survey of these approaches is beyond the scope of this paper, some of the most recent developments can be found in, e.g., \cite{noack2008finite,quarteroni2011certified,wang2012proper} and references therein.
The vast majority of these ROM closure models have been of functional type.
Just as in LES, these functional ROM closure models have generally used some sort of stabilization procedure to model the effect of the discarded modes. 
A physical motivation for this popular approach is given in~\cite{CSB03}, where it is shown that the concept of energy cascade is also valid in a POD setting. 
We believe that the ROM closure models are generally of functional type mainly because the concept of spatial filtering has not been exploited in a ROM setting (see, however, \cite{sabetghadam2012alpha,wang2012proper,wells2015regularized} for a few exceptions.)

In this paper, we use {\it explicit ROM spatial filtering} to construct an LES-ROM framework.
In this LES-ROM framework, we use a novel structural ROM closure model based on AD.
Specifically, given the approximation of the filtered ROM variables, we use AD to obtain an approximation of the original unfiltered ROM variables and solve the ROM closure problem.
Since the AD problem is notoriously ill-posed~\cite{bertero1998introduction,layton2012approximate,vogel2002computational}, we use regularization methods from image processing and inverse problems to obtain stable AD approximations.
Note that the resulting new AD-ROM is fundamentally different from the calibration ROMs used in~\cite{alekseev2001analysis,cordier2010calibration,wang20152d,weller2009robust}, which did not use an explicit ROM spatial filter.

The rest of the paper is organized as follows:  
In Section~\ref{sec:rom}, the POD and the standard ROM are briefly discussed, the LES-ROM framework is developed and the new AD-ROM is proposed within this framework.
The ROM spatial filter (i.e., the ROM differential filter) is discussed in Section~\ref{sec:rom-spatial-filtering}. 
Section~\ref{sec:rom-ed} shows that the exact deconvolution is ill-posed in a ROM context.
Thus, the AD procedure in the AD-ROM needs to employ regularization methods, which are presented in Section~\ref{sec:regularization-methods}.
Section~\ref{sec:rom-ad} shows that the ROM approximate deconvolution using regularization methods is significantly more accurate than the ROM exact deconvolution. 
Numerical results for the new AD-ROM are presented in Section~\ref{sec:numerical-results} for the one-dimensional (1D) Burgers equation. 
Finally, conclusions are drawn in Section~\ref{sec:conclusions}.

\section{Approximate Deconvolution ROM}
    \label{sec:rom}

In this section, we develop the new AD-ROM in several stages.
First, we briefly present the POD (Section~\ref{sec:pod}) and the standard ROM (Section~\ref{sec:g-rom}).
Then, we develop the LES-ROM framework (Section~\ref{sec:les-rom}), which is centered around ROM spatial filtering (Section~\ref{sec:rom-spatial-filtering}).
Finally, within the LES-ROM framework, we construct the new AD-ROM by using the approximate deconvolution methodology (Section~\ref{sec:ad-rom}).

The NSE are used as mathematical model:
\begin{eqnarray}
    && \frac{\partial \bu}{\partial t}
    - Re^{-1} \Delta \bu
    + \bu \cdot \nabla \bu
    + \nabla p
    = {\bf 0} \, ,
    \label{eqn:nse-1}                                                         \\
    && \nabla \cdot \bu
    = 0 \, ,
    \label{eqn:nse-2}
\end{eqnarray}
where $\bu$ is the velocity, $p$ the pressure and $Re$ the Reynolds number. 
The NSE~\eqref{eqn:nse-1}--\eqref{eqn:nse-2} are supplemented with the initial condition $\bu(\bx, 0) = \bu_0(\bx)$ and homogeneous Dirichlet boundary condition: $\bu(\bx, t) = {\bf 0}$ on the boundary.

    \subsection{Proper Orthogonal Decomposition}
        \label{sec:pod}
        One of the most popular reduced order modeling techniques is the
        POD~\cite{HLB96,noack2011reduced,Sir87abc}. For the snapshots
        $\{\bu^1_h,\ldots, \bu^{N_s}_h\}$, which are, e.g., finite element (FE) solutions of
        \eqref{eqn:nse-1}--\eqref{eqn:nse-2} at $N_s$ different time instances,
        the POD seeks a low-dimensional basis that approximates the snapshots
        optimally with respect to a certain norm. In this paper, the commonly
        used $L^2(\Omega)$-norm will be chosen. The solution of the minimization
        problem is equivalent to the solution of the eigenvalue problem
        \begin{equation}
            \label{eq:pod_ev}
            YY^TM \bphi_j = \lambda_j \bphi_j,
            \quad j=1,\ldots,N,
        \end{equation}
        where $\bphi_j$ and $\lambda_j$ denote the vector of the FE coefficients
        of the POD basis functions and the POD eigenvalues, respectively, $Y$
        denotes the snapshot matrix, whose columns correspond to the FE
        coefficients of the snapshots, $M$ denotes the FE mass matrix, and $N$
        is the dimension of the FE space $\bX^h$.
        The eigenvalues are real and non-negative, so they can be ordered as
        follows:$\lambda_1 \ge \lambda_2 \ge \ldots \ge \lambda_R \ge \lambda_{R + 1} = \ldots = \lambda_N = 0$.
        The POD basis consists of the normalized functions $\{
        \bphi_{j}\}_{j=1}^{r}$, which correspond to the first $r\le N$ largest
        eigenvalues. Thus, the POD space is defined as $\bX^r := \text{span} \{
        \bphi_1, \ldots, \bphi_r \}$.

\subsection{Galerkin ROM (G-ROM)}
    \label{sec:g-rom}
    The POD approximation of the velocity is ${\bu}_r({\bf x},t) \equiv
    \sum_{j=1}^r a_j(t) \bphi_j({\bf x})$, where $\{a_{j}(t)\}_{j=1}^{r}$ are
    the sought time-varying coefficients that represent the POD-Galerkin
    trajectories. 
Replacing the velocity $\bu$ by $\bu_r$ in the NSE~\eqref{eqn:nse-1}, we obtain
\begin{eqnarray}
    && \frac{\partial \bu_r}{\partial t}
    - Re^{-1} \Delta \bu_r
    + \bu_r \cdot \nabla \bu_r
    + \nabla p
    = 0 \, ,
    \label{eqn:nse-rom}
\end{eqnarray}
Using a Galerkin projection of~\eqref{eqn:nse-rom} onto $\bX^r$, the {\it Galerkin ROM (G-ROM)} is obtained: $\forall \, k = 1, \ldots, r,$
    \begin{eqnarray}
        \left(
            \frac{\partial \bu_r}{\partial t} , \bphi_{k}
        \right)
        + Re^{-1} \, \left(
            \nabla \bu_r ,
            \nabla \bphi_{k}
        \right)
        + \biggl(
            (\bu_r \cdot \nabla) \bu_r ,
            \bphi_{k}
        \biggr)
        = 0 \, .
    \label{eqn:g-rom}
    \end{eqnarray}

Since $r$ is usually small (e.g., $r= \mathcal{O}(10)$), the G-ROM provides an efficient surrogate model for simulating laminar flows. 
However, for high $Re$ flows, the G-ROM is not a viable tool~\cite{AHLS88, balajewicz2013low, ballarin2015supremizer,quarteroni2011certified, wang2012proper}. 
The main reason for the G-ROM's poor performance in high $Re$ flows is that the POD modes $\{ \bphi_j \}_{j=r+1}^{R}$, which are not used in the G-ROM, play an important role in the physical evolution of the system. 
Thus, the {\it ROM closure problem}~\cite{wang2012proper} needs to be addressed, i.e., the ROM needs to model the effect of the discarded POD modes $\{ \bphi_j \}_{j=r+1}^{R}$. 
If the ROM closure problem is not addressed, then the standard G-ROM generally yields inaccurate results for complex flows, often in the form of numerical oscillations. 

\subsection{ Large Eddy Simulation ROMs (LES-ROMs)}
	\label{sec:les-rom}

Numerous ROM closure modeling approaches have been proposed to address the inaccuracy (and numerical instability) of the standard G-ROM (see, e.g., \cite{noack2011reduced, quarteroni2011certified,wang2012proper} and references therein).
In this paper, we take a different approach based on LES methodology.
The LES-ROM framework is centered around ROM spatial filtering, which is presented in detail in Section~\ref{sec:rom-spatial-filtering}.
To the best of our knowledge, the LES-ROM framework has only been used in Section 3.3.4 in~\cite{wang2012proper}, in the definition of the dynamic sub grid-scale ROM.
Since the LES-ROM framework is essential to the AD-ROM development, we briefly present its main components next.

Using the fact that $\nabla \cdot \bu_r = 0$ in~\eqref{eqn:nse-rom}, we get 
$( \bu_r \cdot \nabla) \, \bu_r = \nabla \cdot (\bu_r \, \bu_r)$.
Thus, \eqref{eqn:nse-rom} can be rewritten as
\begin{eqnarray}
\frac{\partial \bu_r}{\partial t}
- Re^{-1} \Delta \bu_r
+ \nabla \cdot (\bu_r \, \bu_r)
+ \nabla p
= 0 .
\label{eqn:les-rom-1}
\end{eqnarray}
Just as in LES, we filter all the terms in~\eqref{eqn:les-rom-1} with a ROM spatial filter (which will be defined explicitly in Section~\ref{sec:rom-spatial-filtering}).
We then use the fact that the ROM spatial filter is a linear operator and assume that differentiation and ROM filtering commute. 
We obtain 
\begin{eqnarray}
\frac{\partial \obu_r}{\partial t}
- Re^{-1} \Delta \obu_r
+ \nabla \cdot ({\overline{\bu_r \, \bu_r}})
+ \nabla \op
= 0 .
\label{eqn:les-rom-2}
\end{eqnarray}
\begin{remark}[ROM Commutation Error]
If filtering and differentiation do not commute, one has to estimate the commutation error~\cite{BIL05}.
\label{remark:commutation-error}
\end{remark}
Equation~\eqref{eqn:les-rom-2} can be rewritten as
\begin{eqnarray}
\frac{\partial \obu_r}{\partial t}
- Re^{-1} \Delta \obu_r
+ \nabla \cdot (\obu_r \, \obu_r)
+ \nabla \cdot \btau_r
+ \nabla \op
= 0 ,
\label{eqn:les-rom-3}
\end{eqnarray}
where
\begin{eqnarray}
\btau_r
= {\overline{\bu_r \, \bu_r}} - \obu_r \, \obu_r
\label{eqn:les-rom-4}
\end{eqnarray}
is the {\it ROM subfilter-scale stress tensor}.
Using a Galerkin projection of~\eqref{eqn:les-rom-4} onto $\bX^r$ and the fact that $\nabla \cdot \bu_r = 0$, the {\it spatially filtered G-ROM} is obtained: $\forall \, k = 1, \ldots, r,$
    \begin{eqnarray}
        \left(
            \frac{\partial \obu_r}{\partial t} , \bphi_{k}
        \right)
        + Re^{-1} \, \left(
            \nabla \obu_r ,
            \nabla \bphi_{k}
        \right)
        + \biggl(
            (\obu_r \cdot \nabla) \obu_r ,
            \bphi_{k}
        \biggr)
        + \left(
            \btau_r ,
            \nabla \bphi_{k}
        \right)
        = 0 \, .
    \label{eqn:les-rom-5}
    \end{eqnarray}
The spatial structures in the spatially filtered G-ROM~\eqref{eqn:les-rom-5} are larger than the spatial structures in the G-ROM~\eqref{eqn:g-rom}.
Thus, it is expected that the spatially filtered G-ROM requires fewer POD modes than the G-ROM, which is advantageous from a computational point of view.

Of course, the spatially filtered G-ROM~\eqref{eqn:les-rom-5} is not closed.
Thus, just as in LES, one needs to address the ROM closure problem, i.e., to model the ROM subfilter-scale stress tensor $\btau_r$ in terms of the ROM filtered velocity $\obu_r$.
Once the ROM closure problem is addressed, the resulting ROM is called {\it large eddy simulation ROM (LES-ROM)}.

\subsection{Approximate Deconvolution ROM (AD-ROM)}
	\label{sec:ad-rom}
	
Most ROM closure models have generally used some sort of stabilization procedure~\cite{noack2008finite,quarteroni2011certified,wang2012proper}.	
A physical motivation for this popular approach is given in~\cite{CSB03}, where it is shown that the concept of energy cascade is also valid in a POD setting. 
Thus, the role of the neglected POD modes is to drain energy out of the system.
In this paper, we use a completely different approach, based on AD.

The deconvolution idea, which is central in {\it image processing} and {\it inverse problems}~\cite{bertero1998introduction,hansen2010discrete,vogel2002computational}, is simple:
Given an approximation of the filtered input signal, find an approximation of the input itself.
Specifically, in our ROM setting, denoting the spatial filter operator by $G$, we assume that an approximation of the filtered flow variable
\begin{equation}
	\obu_r
	:= G \, \bu_r
	\label{eqn:ad-rom-1}
\end{equation}
is available.
The goal in the deconvolution problem is to find the original flow variable, $\bu_r$.
Since the DF presented in Section~\ref{sec:rom-spatial-filtering} is invertible, at first glance, one just has to use the inverse of the filter $G$ in~\eqref{eqn:ad-rom-1} to solve the deconvolution problem:
\begin{equation}
	{\bu_r}^{ED}
	= G^{-1} \, \obu_r \, .
	\label{eqn:ad-rom-2}
\end{equation}
In inverse problems, this is generally a bad idea because the inverse problem~\eqref{eqn:ad-rom-1} is {\it ill-posed}: small changes in the data can lead to large changes in the solution.
Indeed, inverting the operator $G$ in~\eqref{eqn:ad-rom-2} results in division by small coefficients of the high-frequency components of the operator $G$~\cite{bertero1998introduction,layton2012approximate,vogel2002computational}.
Thus, any changes in the input~\eqref{eqn:ad-rom-1} translate into large nonphysical oscillations in the output~\eqref{eqn:ad-rom-2}~\cite{bertero1998introduction,vogel2002computational}. 
The fact that~\eqref{eqn:ad-rom-2} is useless in our LES-ROM framework is carefully shown in Section~\ref{sec:rom-ed}.

To find a useful approximation to the ROM deconvolution problem~\eqref{eqn:ad-rom-1}, in Section~\ref{sec:rom-ad} we use some of the regularization methods developed in the image processing and inverse problems communities (which are outlined in Section~\ref{sec:regularization-methods}) to obtain an AD approximation:
\begin{equation}
	{\bu_r}^{AD}
	\approx {\bu_r}^{ED}
	= G^{-1} \, \obu_r \, .
	\label{eqn:ad-rom-3}
\end{equation}
The AD approximation~\eqref{eqn:ad-rom-3} is then used to close the spatially filtered G-ROM~\eqref{eqn:les-rom-5}: 
    \begin{eqnarray}
       &&  \left(
            \frac{\partial \obu_r}{\partial t} , \bphi_{k}
        \right)
        + Re^{-1} \, \left(
            \nabla \obu_r ,
            \nabla \bphi_{k}
        \right)
        + \biggl(
            {\overline{\bu^{AD}_r \, \bu^{AD}_r}} ,
            \nabla \bphi_{k}
        \biggr) 
        = 0 \, .
    \label{eqn:ad-rom-4}
    \end{eqnarray}
Thus, we obtain a new LES-ROM, the {\it approximate deconvolution ROM (AD-ROM)}:
    \begin{eqnarray}
       &&  \left(
            \frac{\partial \bw_r}{\partial t} , \bphi_{k}
        \right)
        + Re^{-1} \, \left(
            \nabla \bw_r ,
            \nabla \bphi_{k}
        \right)
        + \biggl(
            {\overline{\bw^{AD}_r \, \bw^{AD}_r}} ,
            \nabla \bphi_{k}
        \biggr) 
        = 0 \, ,
    \label{eqn:ad-rom}
    \end{eqnarray}
where $\bw_r$ is an approximation of the spatially filtered flow variable $\obu_r$.

\section{ROM Spatial Filtering}
    \label{sec:rom-spatial-filtering}
    
To develop the LES-ROM framework, and in particular the new AD-ROM~\eqref{eqn:ad-rom}, we used a generic ROM spatial filter.
In this section we describe the ROM spatial filter that is used in the numerical investigation in Section~\ref{sec:numerical-results}.

In standard reduced order modeling, spatial filtering is generally defined implicitly, by truncating the POD basis used in the Galerkin approximation. 
To our knowledge, explicit ROM spatial filtering has only be used in~\cite{wang2012proper,sabetghadam2012alpha,wells2015regularized}.
Two types of explicit ROM spatial filters were used in these papers: a ROM projection and a ROM differential filter.
In this paper, we exclusively use the ROM differential filter, since it yielded better results than the ROM projection in the numerical investigation of regularized ROMs~\cite{wells2015regularized}.
We emphasize, however, that other ROM spatial filters could be used equally well in the new AD-ROM~\eqref{eqn:ad-rom}.

The ROM differential filter is defined as follows: 
Let $\delta$ be the radius of the ROM differential filter, which determines the length scales modeled in the LES-ROM framework.
For a given $\bu^r \in \bX^r$, the {\it ROM differential filter (DF)} seeks $\obu^r \in \bX^r$ such that
\begin{eqnarray}
		\biggl(
			\left( I - \delta^2 \Delta \right) \obu^r , \bphi_j
          \biggr)
          = (\bu^r, \bphi_j),
           \quad \forall \, j=1, \ldots r \, .
	\label{eqn:df}
\end{eqnarray}
The differential filter has been introduced by Germano in LES~\cite{germano1986differential-b,germano1986differential}. 
It was also used in a ROM context to develop regularized ROMs: 
The DF was first used in~\cite{sabetghadam2012alpha} for the 1D Kuramoto--Sivashinsky equation in a periodic setting.
The DF was subsequently used in~\cite{wells2015regularized} for the 3D NSE in a general non-periodic setting.

The DF~\eqref{eqn:df} has several appealing properties~\cite{BIL05}:
First, it acts as a spatial filter, since it eliminates the small scales (i.e., high frequencies) from the input. 
Indeed, the DF uses an elliptic operator to smooth the input variable.     
Second, the DF has an extremely low computational overhead, since it amounts to solving a linear system with a very small $r \times r$ matrix that is precomputed.
Third, the DF preserves incompressibility in the NSE, since the POD basis functions are incompressible and the DF is a linear operator.

The algorithmic decisions needed in the implementation of the DF are carefully discussed in Section 3.4 in~\cite{wells2015regularized}.
Next, we summarize these decisions.
If the input variable is in $\bX^r$, i.e., $\bu = \sum_{j=1}^{r} U_j \, \bphi_j$, then the following linear system needs to be solved:
\begin{eqnarray}
	\left( \bM + \delta^2 \, \bS \right) \, \obU
	= \bU \, ,
	\label{eqn:pod-df-linear-system}
\end{eqnarray}
where 
$\bM \in \R^{r \times r}$ is the POD mass matrix with entries $(\bS)_{i j} = ( \bphi_j , \bphi_i), \, 1 \leq i, j \leq r$,
$\bS \in \R^{r \times r}$ is the POD stiffness matrix with entries $(\bS)_{i j} = (\nabla \bphi_j , \nabla \bphi_i)$,  $1 \leq i, j \leq r$,
$\obU \in \R^{r}$ is the vector of coefficients of the output filtered variable $\obu$ with entries $(\obU)_{i} = \oU_i, \, 1 \leq i \leq r$, and
$\bU \in \R^{r}$ is the vector with entries $(\bU)_{i} = (\bu , \bphi_i), \, 1 \leq i \leq r$.
If the input variable is not in $\bX^r$ (e.g., the input variable is the centering trajectory), then the finite element version of the DF is used to prefilter the input variable (see Section 3.4 in~\cite{wells2015regularized} for details).

\section{ROM Exact Deconvolution}
\label{sec:rom-ed}

In this section, we answer the following question: 
{\it Is the ROM exact deconvolution problem ill-posed?}
That is, given a signal filtered with the DF~\eqref{eqn:df}, can the original signal be recovered by applying the inverse of the DF to the filtered signal?
In our ROM setting, this would translate in investigating whether the ROM exact deconvolution problem~\eqref{eqn:ad-rom-2} is both stable and accurate.

We emphasize that this is an important practical question.
Indeed, if the ROM exact deconvolution problem~\eqref{eqn:ad-rom-2} is well-posed, then we can simply use the ROM exact deconvolution in the AD-ROM~\eqref{eqn:ad-rom}.
Otherwise, we need to use AD techniques, just as in LES.

Also note that the deconvolution problem was shown to be ill-posed in the LES context~\cite{layton2012approximate}.
We emphasize, however, that since the LES and ROM settings are significantly different, (e.g., the problem sizes are vastly different -- $\mathcal{O}(10^6)$ for LES and $\mathcal{O}(10)$ for ROM), we need to explicitly investigate whether the ROM exact deconvolution problem is ill-posed.
To this end, we consider the following algorithm:

\begin{center}
	{\bf Algorithm 1}
\end{center}
\begin{itemize}
	\item[(1)] Consider input signal $u$.
	\item[(2)] Calculate the filtered input signal $\ou := G \, u$, where $G$ is the DF~\eqref{eqn:df}.
	\item[(3)] {\it Exactly deconvolve (ED)} the input signal:
				\begin{equation} 
					u^{ED} 
					= G^{-1} \, \ou 
					= ({\bf M} + \delta^2 \, {\bf S}) \, \ou \, .
					\label{eqn:ed}
				\end{equation}
	\item[(4)] Compare the ED signal $u^{ED}$ with the true signal $u$.
\end{itemize}

One of the following two cases can result from Algorithm 1:
{\it Case 1}:  $u^{ED}$ is close to $u$. 
In this case, we use the ED method~\eqref{eqn:ed} in the AD-ROM~\eqref{eqn:ad-rom}, i.e., we solve small linear systems without any regularization.
{\it Case 2}:   $u^{ED}$ is not close to $u$.
In this case, we need to use the regularization methods presented in Section~\ref{sec:regularization-methods} in the AD-ROM~\eqref{eqn:ad-rom}.

\paragraph{Noisy Data}
In the numerical investigation in this section, the effect of {\it noise} on the filtered input signal will play a central role.
Specifically, Algorithm 1 will be modified as follows.

\clearpage

\begin{center}
	{\bf Algorithm 1'}
\end{center}
\begin{itemize}
	\item[(1)] Consider input signal $u$.
	\item[(2)] Calculate the filtered input signal $\ou := G \, u$, where $G$ is the DF~\eqref{eqn:df}.
	\item[(3)] {\it Exactly deconvolve (ED)} the {\it noisy} input signal:
				\begin{equation} 
					u^{ED} 
					= G^{-1} \, (\ou + \eta) 
					= ({\bf M} + \delta^2 \, {\bf S}) \, (\ou + \eta) \, ,
					\label{eqn:ed-noise}
				\end{equation}
				where $\eta$ is the noise.
	\item[(4)] Compare the ED signal $u^{ED}$ with the true signal $u$.
\end{itemize}

Adding noise to the filtered input signal is common practice in image processing and inverse problems~\cite{bertero1998introduction,hansen2010discrete,vogel2002computational}.
This is also relevant to our AD-ROM~\eqref{eqn:ad-rom}, where noise could model, e.g., the inherent error in the numerical approximation of the filtered flow variable, $\obu_r$.
Indeed, in the numerical discretization of the AD-ROM~\eqref{eqn:ad-rom}, the filtered flow variable $\obu_r$ is generally not available.
Instead, a {\it numerical approximation} of the filtered flow variable is available.
To further reinforce this point that the added noise models the AD-ROM numerical discretization error, the magnitude of the noise used in this section is within the magnitude of the AD-ROM's spatial discretization error (see Remark~\ref{remark:noise}).

\paragraph{Condition Number}
Another important role in ill-posed inverse problems is played by the condition number of the problem~~\cite{bertero1998introduction,hansen2010discrete,vogel2002computational}.
Indeed, if the condition number is large, then the noise in the input data will generally be highly amplified in the deconvolution process.
Thus, in our setting, the condition number of the matrix ${\bf M} + \delta^2 \, {\bf S}$ of the linear system~\eqref{eqn:ed-noise} needs to be investigated.

From a theoretical point of view, estimating the condition number of ${\bf M} + \delta^2 \, {\bf S}$ is generally impossible, since this matrix depends on the particular POD basis functions used, which in turn depend on the particular physical setting employed.
To get some insight into this condition number, however, we use the approach proposed in~\cite{iliescu2014are} and consider the case when the POD basis is actually the Fourier basis.
(This is the case when the underlying problem is homogeneous~\cite{HLB96}.)
As shown in~\cite{iliescu2014are}, in this case, $\bM = \frac{1}{2} \, {\bf I}$. 
Furthermore, since the POD basis functions are $\varphi_j(x) = \sin(j \, \pi \, x)$ (i.e., the Fourier basis functions), the matrix $\bS$ is diagonal and its diagonal entries are given by $S_{j j} = \frac{1}{2} (j \, \pi)^2$.
Finally, since the matrix $({\bf M} + \delta^2 \, {\bf S})$ is symmetric, its matrix 2-norm is given by the largest eigenvalue.
Thus, the condition number of the ROM exact deconvolution, $\mK^{ED}$, can be computed as follows:
\begin{equation}
	\mK^{ED}
	:= \mK\bigl( ({\bf M} + \delta^2 \, {\bf S})^{-1} \bigr)
	= \| ({\bf M} + \delta^2 \, {\bf S})^{-1} \|_2  \, \| ({\bf M} + \delta^2 \, {\bf S}) \|_2 
	= \frac{ \frac{1}{2} + \frac{1}{2} \, \delta^2 \, (r \, \pi)^2 }{\frac{1}{2} + \frac{1}{2} \, \delta^2 \, (1 \, \pi)^2} \, .
	\label{eqn:rom-ad-ill-posed-1}
\end{equation}
The condition number in~\eqref{eqn:rom-ad-ill-posed-1} scales asymptotically as follows:
\begin{equation}
	\mK^{ED} \sim
		\begin{cases}
   			r^2 & \text{if } \delta^2 r^2 \text{ large} \\
   			1    & \text{if } \delta^2 r^2 \text{ small} \, .
  		\end{cases}
	\label{eqn:rom-ad-ill-posed-2}
\end{equation}

\paragraph{Numerical Results}
To investigate whether the ROM exact deconvolution is ill-posed, we use Algorithm 1 (in which noise is not considered) and Algorithm 1' (in which noise is considered).
The following parameter choices are made in these two algorithms: 
The input signal is 
$u(x) = \sin(2 \, \pi \, x)  + 0.1 \sin(4 \, \pi \, x) + 0.1 \sin(16 \, \pi \, x)$ and the POD basis is the Fourier basis $\{ \sin(\pi \, x), \sin(2 \, \pi \, x), \ldots \sin(16 \, \pi \, x) \}$.
Furthermore, random noises of three different magnitude orders are considered: $\mathcal{O}(10^{-2})$, $\mathcal{O}(10^{-3})$, $\mathcal{O}(10^{-4})$; the resulting approximations use subscripts $n_1$, $n_2$ and $n_3$, respectively.
Specifically, the Matlab command ``rand" is first used to generate numbers between $0$ and $1$; these numbers are then divided by $10^2, 10^3$ and $10^4$.

\begin{figure}[h]
\minipage{0.33\textwidth}
  \includegraphics[width=\linewidth]{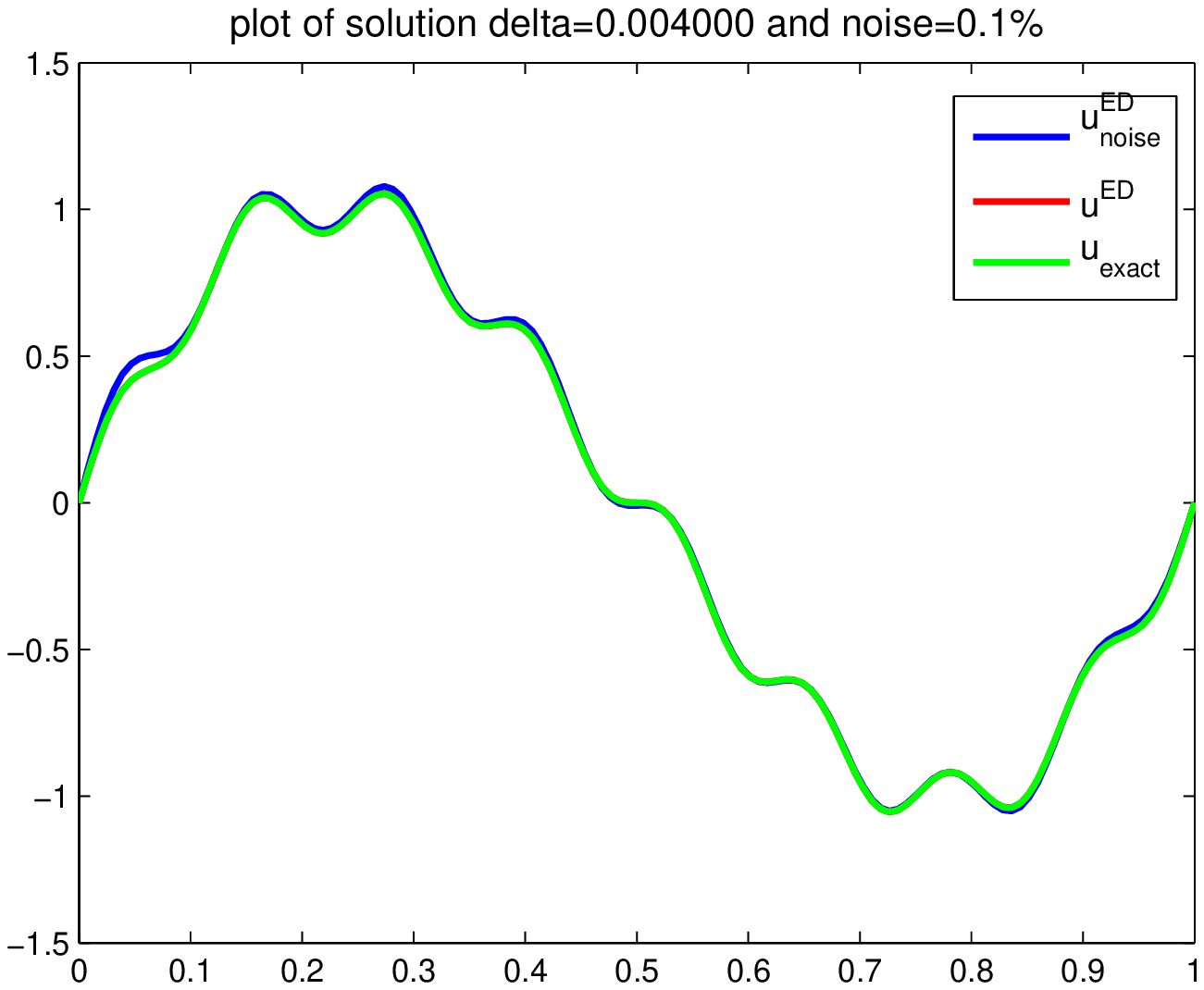}
\endminipage\hfill
\minipage{0.33\textwidth}
  \includegraphics[width=\linewidth]{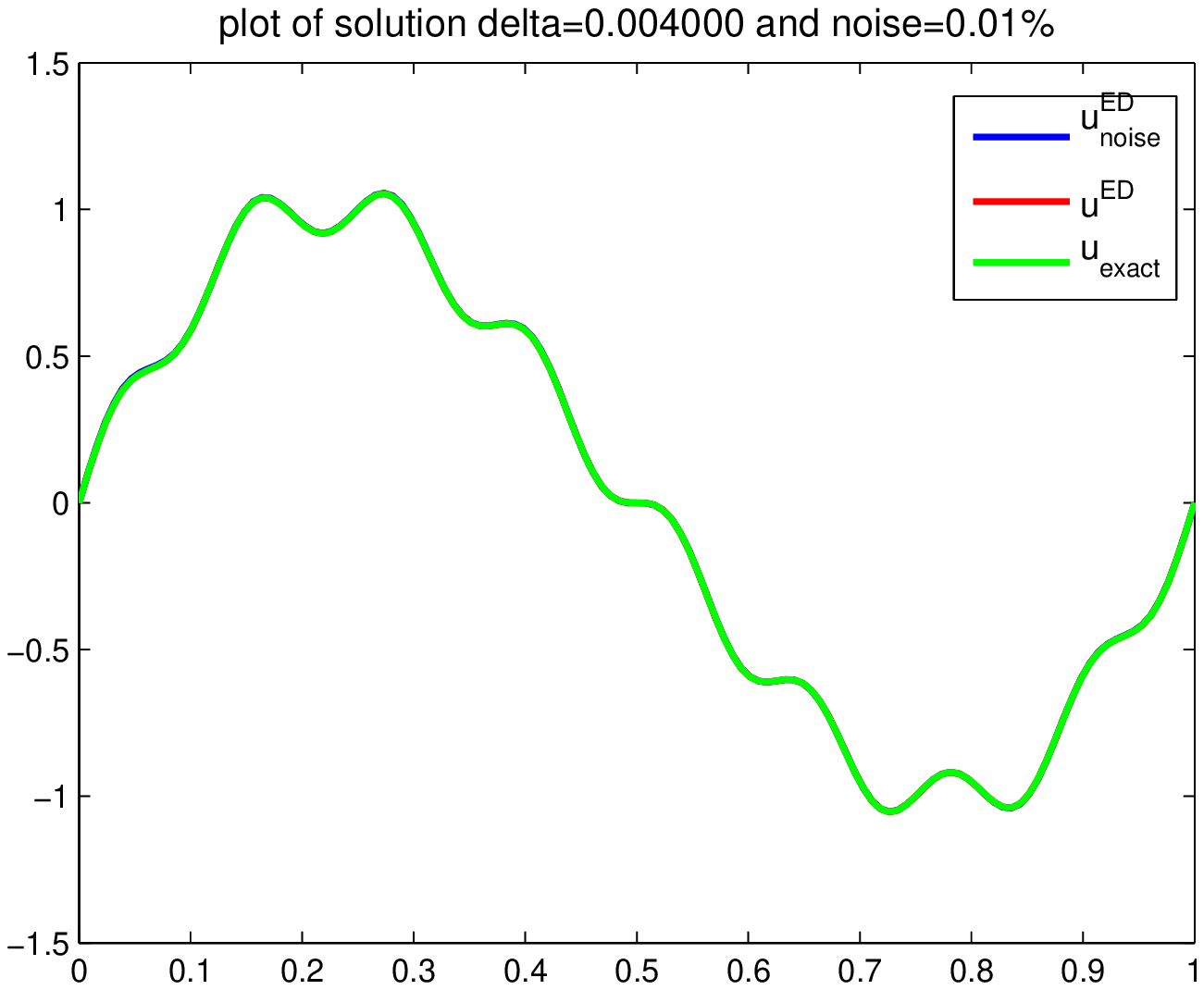}
\endminipage\hfill
\minipage{0.33\textwidth}
  \includegraphics[width=\linewidth]{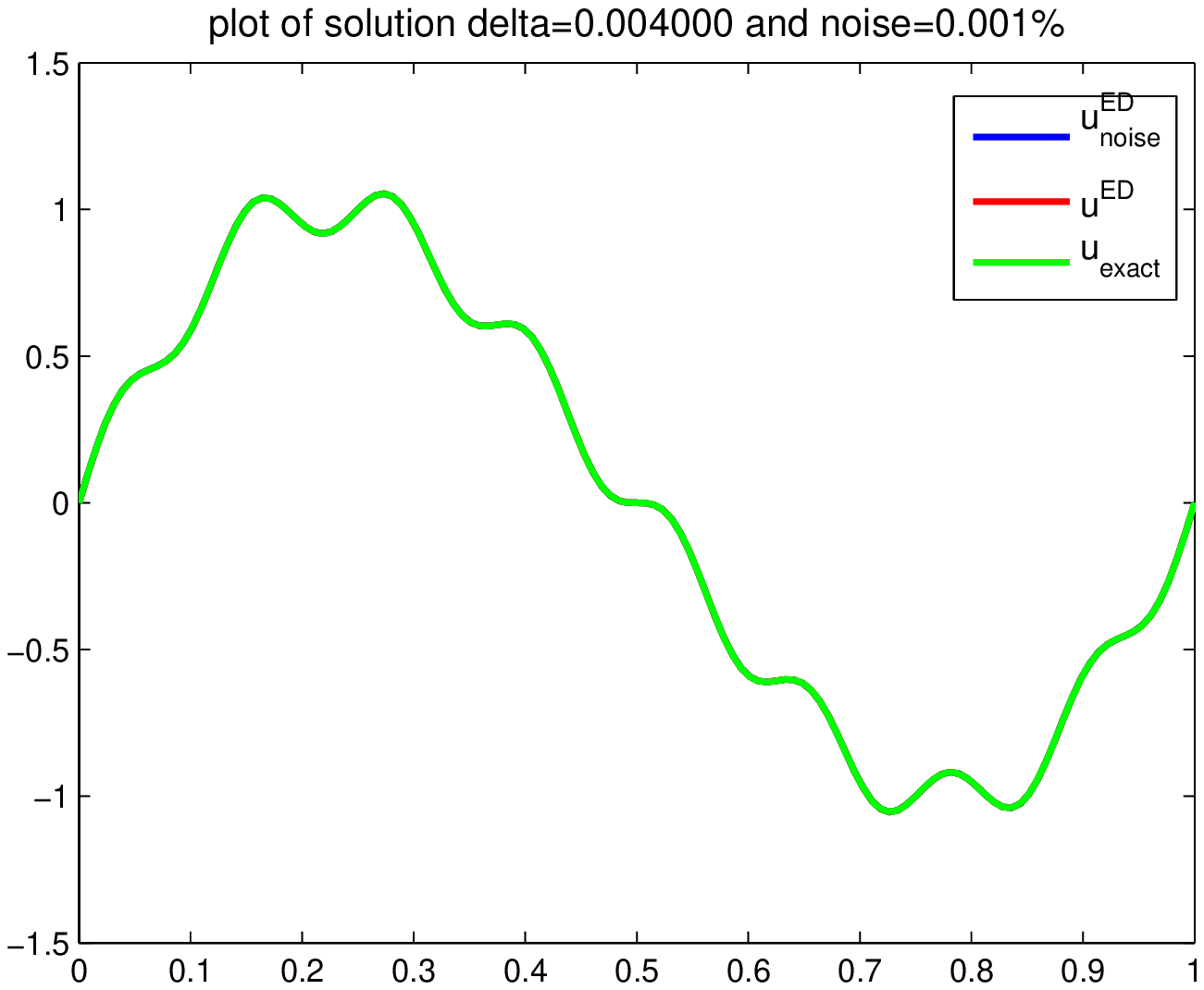}
\endminipage\hfill \\
\minipage{0.33\textwidth}
  \includegraphics[width=\linewidth]{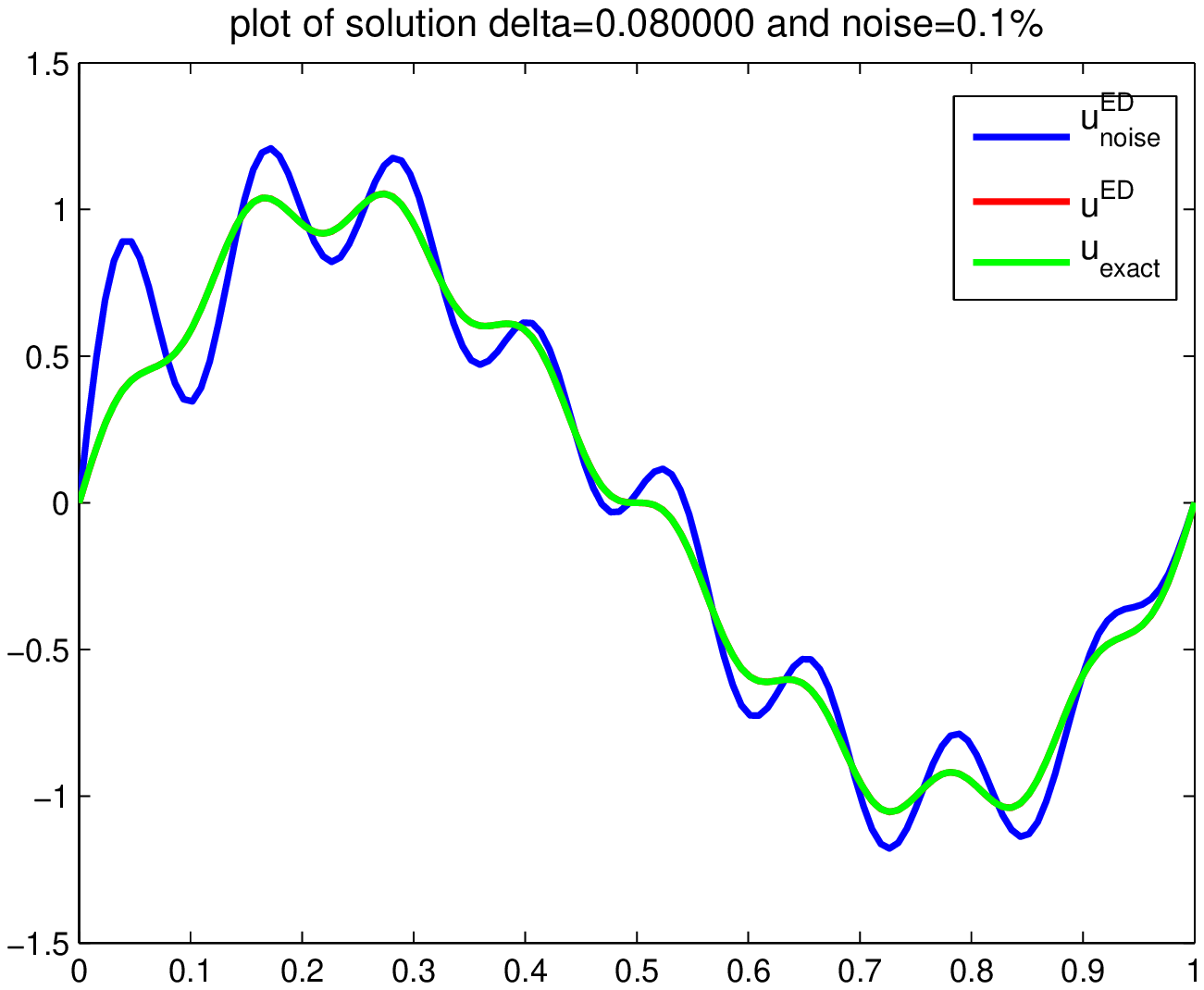}
\endminipage\hfill
\minipage{0.33\textwidth}
  \includegraphics[width=\linewidth]{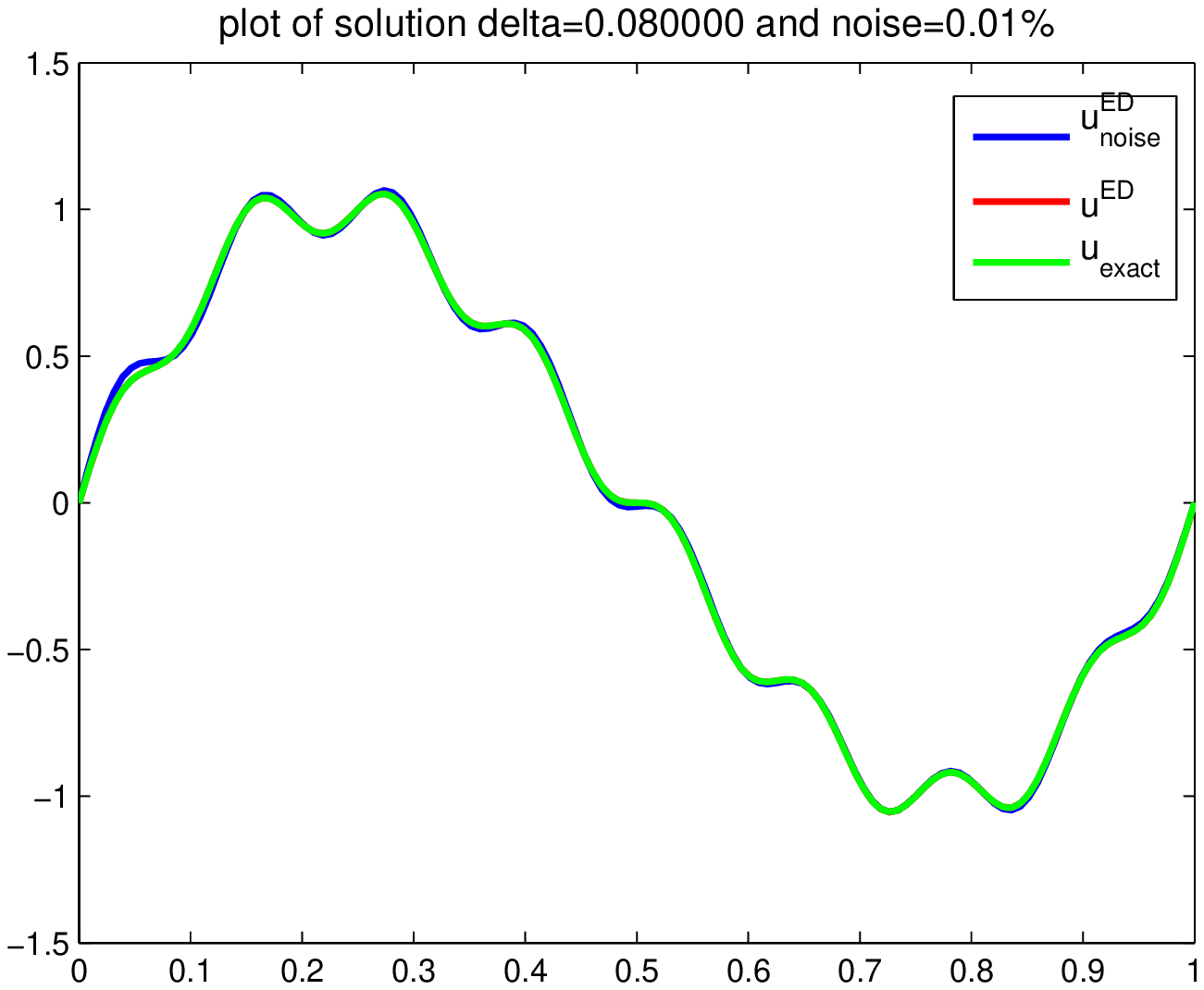}
\endminipage\hfill
\minipage{0.33\textwidth}
  \includegraphics[width=\linewidth]{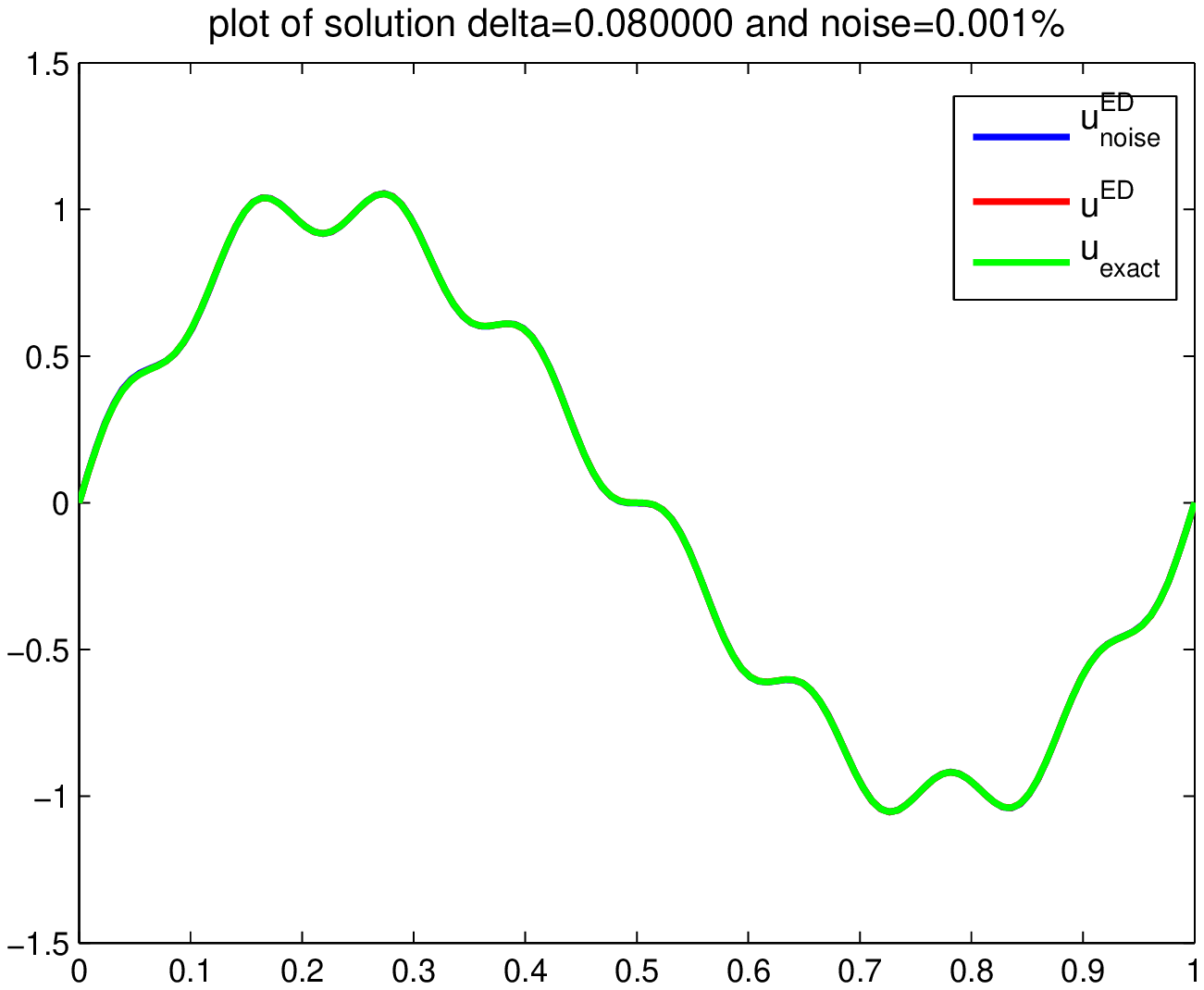}
\endminipage\hfill \\
\minipage{0.33\textwidth}
  \includegraphics[width=\linewidth]{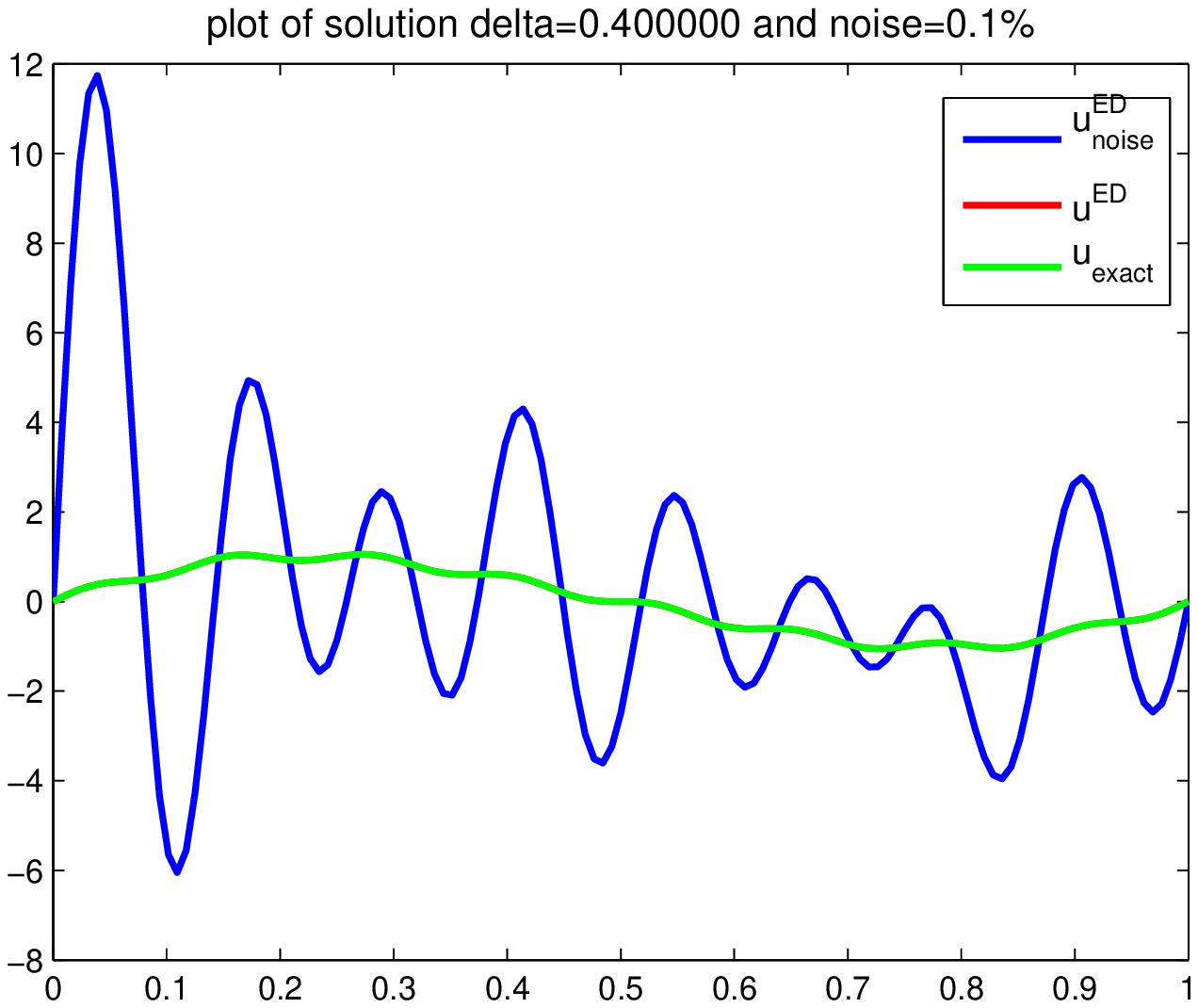}
\endminipage\hfill
\minipage{0.33\textwidth}
  \includegraphics[width=\linewidth]{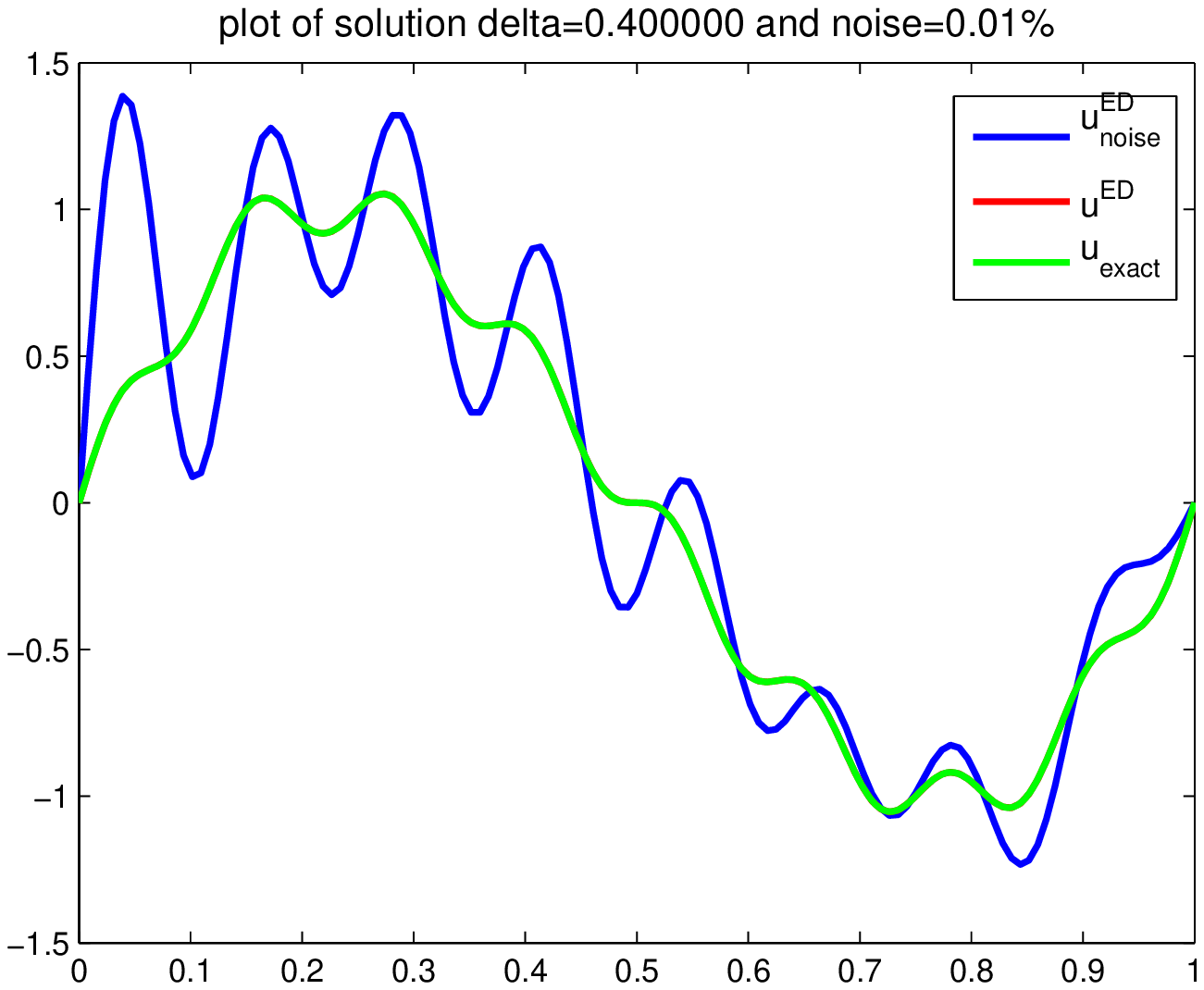}
\endminipage\hfill
\minipage{0.33\textwidth}
  \includegraphics[width=\linewidth]{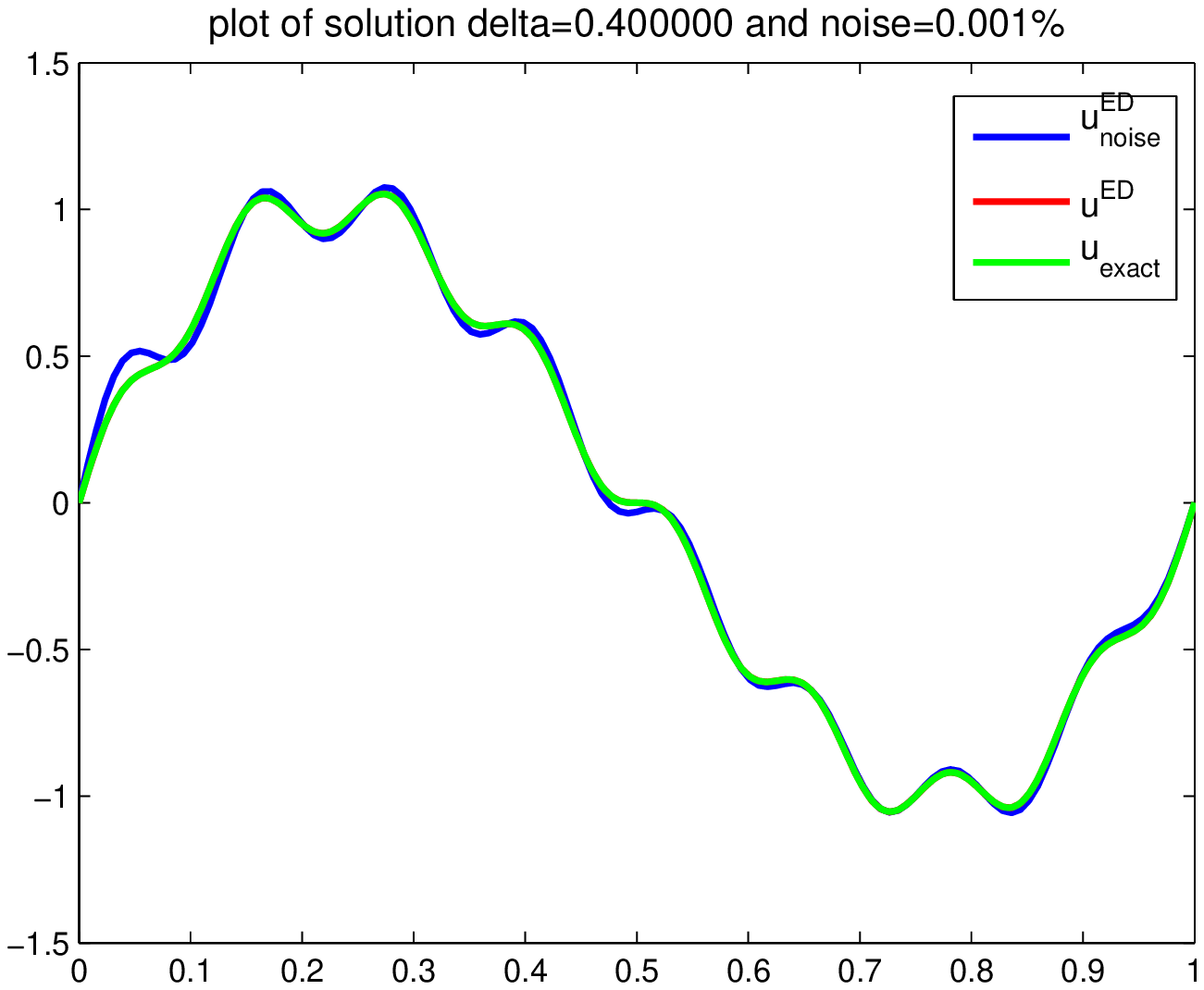}
\endminipage\hfill
	\caption{
		ROM exact deconvolution.
		Plots of the approximations without noise, i.e., with Algorithm 1 (red curve) and with noise, i.e., with Algorithm 1' (blue curve). 
		Three different $\delta$ values are used:
		$\delta = 0.004$ (top row), $\delta = 0.08$ (middle row) and $\delta = 0.4$ (bottom row).
		Three different noise magnitude levels are used: 
		$\mathcal{O}(10^{-3})$ (left column), $\mathcal{O}(10^{-4})$ (middle column) and $\mathcal{O}(10^{-5})$ (right column).
		The exact solution (green curve) is also plotted for comparison purposes.
		\label{fig:rom-ed-1}
	}
\end{figure}

The ROM exact deconvolution approximation without noise ($u^{ED}$) and the ROM exact deconvolution with the three noise levels ($u^{ED}_{n_1}$, $u^{ED}_{n_2}$ and $u^{ED}_{n_3}$) are plotted in Fig.~\ref{fig:rom-ed-1} for three representative $\delta$ values.
As benchmark, the exact input signal ($u$) is also plotted.
These plots yield the following conclusions:
First, without noise, the ROM exact deconvolution approximation is accurate (almost indistinguishable from the exact input signal).
Second, when noise is added, the ROM exact deconvolution is inaccurate, especially for large $\delta$ values and noise levels.
This behavior is natural, since estimate~\eqref{eqn:rom-ad-ill-posed-1} shows that, for the value $r=16$ that was used in this section, the condition number of the ROM exact deconvolution problem increases with respect to $\delta$ (see sixth column of Table~\ref{table:rom-ed-1}).
These conclusions are reinforced by the errors of $u^{ED}$, $u^{ED}_{n_1}$, $u^{ED}_{n_2}$ and $u^{ED}_{n_3}$, which are listed in Table~\ref{table:rom-ed-1}.

Based on the results in Fig.~\ref{fig:rom-ed-1} and Table~\ref{table:rom-ed-1}, we conclude that, when noise is added to the filtered input signal, the ROM exact deconvolution increases its magnitude, which results in an inaccurate, oscillatory approximation of the original unfiltered signal.
This suggests that the ROM exact deconvolution is ill-posed. 
We also emphasize that, in the simplified setting considered in this section, the condition number of the ROM exact deconvolution, $\mK^{ED}$, is relatively low (see Table~\ref{table:rom-ed-1}).
We expect, however, that $\mK^{ED}$ will be significantly larger in realistic fluid flow settings, which are the new AD-ROM's ultimate target.
Thus, in these realistic settings, we expect the ill-posedness of the ROM exact deconvolution to be more clearly displayed.

\begin{table}[h]
\centering
\begin{tabular}{|l|c|c|c|c|l|}
\hline
\multicolumn{2}{|c|}{{ }} & $\mathcal{O}(10^{-3})$ noise & $\mathcal{O}(10^{-4})$ noise & $\mathcal{O}(10^{-5})$ noise &  \\ \hline
      $\delta$        &  $\|u-u^{ED}\|_{L^2}$            & $\|u-u_{n1}^{ED}\|_{L^2}$ & $\|u-u_{n2}^{ED}\|_{L^2}$ & $\|u-u_{n3}^{ED}\|_{L^2}$ & $\mK^{ED}$ \\ \hline
   $0.004$ & 9.1451e-04 & 0.0167 & 0.0021 & 9.7247e-04 & 1.01 \\ \hline
   $0.08$    & 9.1451e-04 & 0.1377 & 0.0126 & 0.0018 & 15.93 \\ \hline
   $0.4$   & 9.1451e-04 & 3.0670 & 0.2743 & 0.0278 & 155.13 \\ \hline
\end{tabular}
	\caption{
		ROM exact deconvolution.
		Errors without noise, i.e., with Algorithm 1 (second column) and with noise, i.e., with Algorithm 1'  (third, fourth and fifth columns). 
		Three different $\delta$ values and three different different noise magnitude levels are used.
		The condition number is also listed (sixth column).
	}
\label{table:rom-ed-1}
\end{table}

\clearpage

\section{Regularization Methods for Ill-Posed Inverse Problems}
\label{sec:regularization-methods}

In Section~\ref{sec:rom-ed} it was shown that the ROM exact deconvolution problem is ill-posed.
In this section, we briefly describe several numerical methods that are commonly used for general ill-posed inverse problems.
These methods will be tested on the ROM exact deconvolution problem~\eqref{eqn:ad-rom-2} in Section~\ref{sec:rom-ad}. 

The ill-posedness of inverse problems is generally manifested in practice in the form of numerical oscillations (which were also observed in the ROM exact deconvolution in Section~\ref{sec:rom-ed}).
Thus, different {\it regularization methods} have been proposed to alleviate the numerical oscillations displayed by ill-posed inverse problems. 
Next, we briefly summarize the most popular regularization methods.
More details can be found in, e.g., ~\cite{bertero1998introduction,hansen2010discrete,vogel2002computational}.

There are two types of regularization methods: {\it variational} regularization methods, in which one solves a minimization problem (see Section~\ref{sec:variational-regularization-methods}), and {\it iterative} regularization methods, in which one uses a fixed-point iteration to approximate the solution (see Section~\ref{sec:iterative-regularization-methods}).

\subsection{Variational Regularization Methods}
\label{sec:variational-regularization-methods}

Among the variational regularization methods, the Tikhonov and Lavrentiev regularization methods are commonly used.

\paragraph{Tikhonov regularization method}
This is one of the most popular regularizations for ill-posed problems~\cite{bertero1998introduction,hansen2010discrete,vogel2002computational}. 
The Tikhonov regularization method minimizes the following functional (see equation (5.6) in~\cite{bertero1998introduction}):
\begin{equation}
	\Phi_{\mu}(u)
	:= \| G \, u - \ou \|^2
	+ \mu \, \| u \|^2 \, ,
	\label{eqn:tikhonov-1}
\end{equation}
which is a linear combination of the discrepancy functional $\| G \, u - \ou \|^2$ and the  energy functional $\| u \|^2$.
Problem~\eqref{eqn:tikhonov-1} aims at minimizing the discrepancy functional such that the solution has a prescribed energy (see Fig. 5.1 in~\cite{bertero1998introduction}).
The Tikhonov regularization~\eqref{eqn:tikhonov-1} is a variational algorithm which is equivalent to the following linear system~\cite{bertero1998introduction,vogel2002computational}:
\begin{equation}
	(G^* \, G \, + \mu \, I) \, u
	= G^* \, \ou \, .
	\label{eqn:tikhonov-2}
\end{equation}
Thus, to solve the minimization problem in~\eqref{eqn:tikhonov-1}, one simply needs to compute
\begin{equation}
	u
	= (G^* \, G \, + \mu \, I)^{-1} G^* \, \ou \, .
	\label{eqn:tikhonov-3}
\end{equation}
The parameter $\mu$ in~\eqref{eqn:tikhonov-1} is a free parameter. 
Several approaches for finding its optimal value are discussed in~\cite{bertero1998introduction,hansen2010discrete,vogel2002computational}.
For large scale problems, instead of solving the linear system~\eqref{eqn:tikhonov-2}, one can instead use an iterative procedure, i.e., the Tikhonov iterative regularization.

\paragraph{Lavrentiev regularization method}
This is the Tikhonov regularization method, but without the adjoint operator $G^*$.
Specifically, the Lavrentiev regularization method replaces the linear system in~\eqref{eqn:tikhonov-2} with the following linear system (see (12.3) in~\cite{bertero1998introduction}):
\begin{equation}
	(G \, + \mu \, I) \, u
	= \ou \, .
	\label{eqn:lavrentiev-1}
\end{equation}

An improvement to the Lavrentiev regularization method is the {\it modified Lavrentiev regularization method}~\cite{layton2012approximate,stanculescu2010numerical}:
\begin{equation}
	((1 - \mu) \, G \, + \mu \, I) \, u
	= \ou \, .
	\label{eqn:modified-lavrentiev-1}
\end{equation}

\subsection{Iterative Regularization Methods}
\label{sec:iterative-regularization-methods}

Among the iterative regularization methods, the Landweber and Van Cittert regularization methods are commonly used.

\paragraph{Landweber iterative method}
This method minimizes the following functional (see Appendix E in~\cite{bertero1998introduction}):
\begin{equation}
	\Phi(u)
	:= \| G \, u - \ou \|^2 \, ,
	\label{eqn:landweber-1}
\end{equation}
i.e., it minimizes the discrepancy functional $\| G \, u - \ou \|^2$, which is equivalent to the following linear system~\cite{bertero1998introduction}:
\begin{equation}
	(G^* \, G ) \, u
	= G^* \, \ou \, .
	\label{eqn:landweber-2}
\end{equation}
For large scale problems, one can instead use the Landweber iterative method
\begin{equation}
	u^{n+1}
	= u^n + \tau \, \bigl[ G^* \, \ou - (G^* \, G) \, u^n \bigr] \, ,
	\label{eqn:landweber-7}
\end{equation}
where $\tau$ is a relaxation parameter and the number of iterations plays the role of a regularization parameter.

\paragraph{Van Cittert iterative method} 
This method is defined as follows~\cite{bertero1998introduction}:
\begin{equation}
	u^{n+1}
	= u^n + \tau \, (\ou - G \, u^n) \, .
	\label{eqn:van-cittert-2}
\end{equation}
The Van Cittert iterative method is the most popular regularization method in AD-LES~\cite{SA99,SAK01a,SAK01b,layton2009chebychev,layton2012approximate}. \\

We emphasize that our LES-ROM framework is different from the standard LES setting:
The dimension of the problem is small for the LES-ROM framework ($\mathcal{O}(10))$, but very large for the LES setting ($\mathcal{O}(10^6))$.
Thus, it is not clear which regularization method is the most appropriate for our LES-ROM framework.
This issue is discussed in Section~\ref{sec:rom-ad}.

\section{ROM Approximate Deconvolution}
\label{sec:rom-ad}

In Section~\ref{sec:rom-ed} it was shown that the ROM exact deconvolution problem~\eqref{eqn:ad-rom-2} is ill-posed. 
Indeed, using Algorithm 1' showed that making small changes in (i.e., adding noise to) the input filtered signal yields nonphysical numerical oscillations.
This behavior, which is typical to ill-posed inverse problems, is generally mitigated by using some of the regularization methods outlined in Section~\ref{sec:regularization-methods}.
In this section, we use these regularization methods to solve the ROM deconvolution problem. 
Following the LES terminology~\cite{BIL05,layton2012approximate}, we call ROM approximate deconvolution the regularized methods applied to the ROM deconvolution problem.
Just as is usually done in ill-posed inverse problems~\cite{bertero1998introduction,vogel2002computational}, we investigate whether Algorithm 2 (presented below) yields more accurate results than Algorithm 1'.

\begin{center}
	{\bf Algorithm 2}
\end{center}
\begin{itemize}
	\item[(1)] Consider input signal $u$.
	\item[(2)] Calculate the filtered input signal $\ou := G \, u$, where $G$ is the DF.
	\item[(3)] Approximately deconvolve the noisy input signal $u^{AD} \approx G^{-1} \, (\ou + \eta)$, where $\eta$ is the noise.
	\item[(4)] Compare the AD signal $u^{AD}$ with the true signal $u$ and the ED signal $u^{ED}$ from Algorithm 1'.
\end{itemize}

As in Section~\ref{sec:rom-ed}, the effect of noise on the filtered input signal is studied in Step 3 of Algorithm 2.
As mentioned before, this is relevant to the numerical discretization of the new AD-ROM~\eqref{eqn:ad-rom}, where noise could model the numerical error in the approximation of the filtered flow variable, $\obu_r$.

Several regularization methods described in Section~\ref{sec:regularization-methods} (Lavrentiev, modified Lavrentiev, and Van Cittert) were investigated numerically in Step 3 of Algorithm 2.
Since they yielded similar qualitative results in our preliminary tests, for clarity, we exclusively consider the Lavrentiev regularization method~\eqref{eqn:lavrentiev-1} to find $u^{AD}$ in Step 3 of Algorithm 2:
\begin{equation}
	(G \, + \mu \, I) \, u^{AD-L}
	= \ou \, ,
	\label{eqn:fourier-rom-ad-lavrentiev-1}
\end{equation}
where $u^{AD-L}$ is the AD approximation of $u$ obtained with the Lavrentiev regularization method.
Since $G$ is the DF~\eqref{eqn:df}, equation~\eqref{eqn:fourier-rom-ad-lavrentiev-1} becomes
\begin{equation}
	\biggl( \bigl(I - \delta^2 \, \Delta \bigr)^{-1}  + \mu \, I \biggr) \, u^{AD-L}
	= \ou \, .
	\label{eqn:fourier-rom-ad-lavrentiev-2}
\end{equation}
Multiplying~\eqref{eqn:fourier-rom-ad-lavrentiev-2} by $\bigl(I - \delta^2 \, \Delta \bigr)$, we get
\begin{equation}
	u^{AD-L}
	+ \mu \, \bigl(I - \delta^2 \, \Delta \bigr) \, u^{AD-L}
	= \bigl(I - \delta^2 \, \Delta \bigr) \, \ou \, ,
	\label{eqn:fourier-rom-ad-lavrentiev-3}
\end{equation}
which in matrix form can be written as
\begin{equation}
	\bigl( \bM + \mu \, \bM + \mu \, \delta^2 \, \bS \bigr) \, u^{AD-L}
	= \bigl( \bM + \delta^2 \, \bS \bigr) \, \ou \, .
	\label{eqn:fourier-rom-ad-lavrentiev-3}
\end{equation}

In Step 4 of Algorithm 2, the AD signal ($u^{AD-L}$) is compared with the true signal ($u$) and the ED signal ($u^{ED}$) from Step 3 of Algorithm 1'.
We expect that the regularization used in the computation of $u^{AD-L}$ will make it more accurate than $u^{ED}$.

\begin{figure}[h]
\minipage{0.33\textwidth}
  \includegraphics[width=\linewidth]{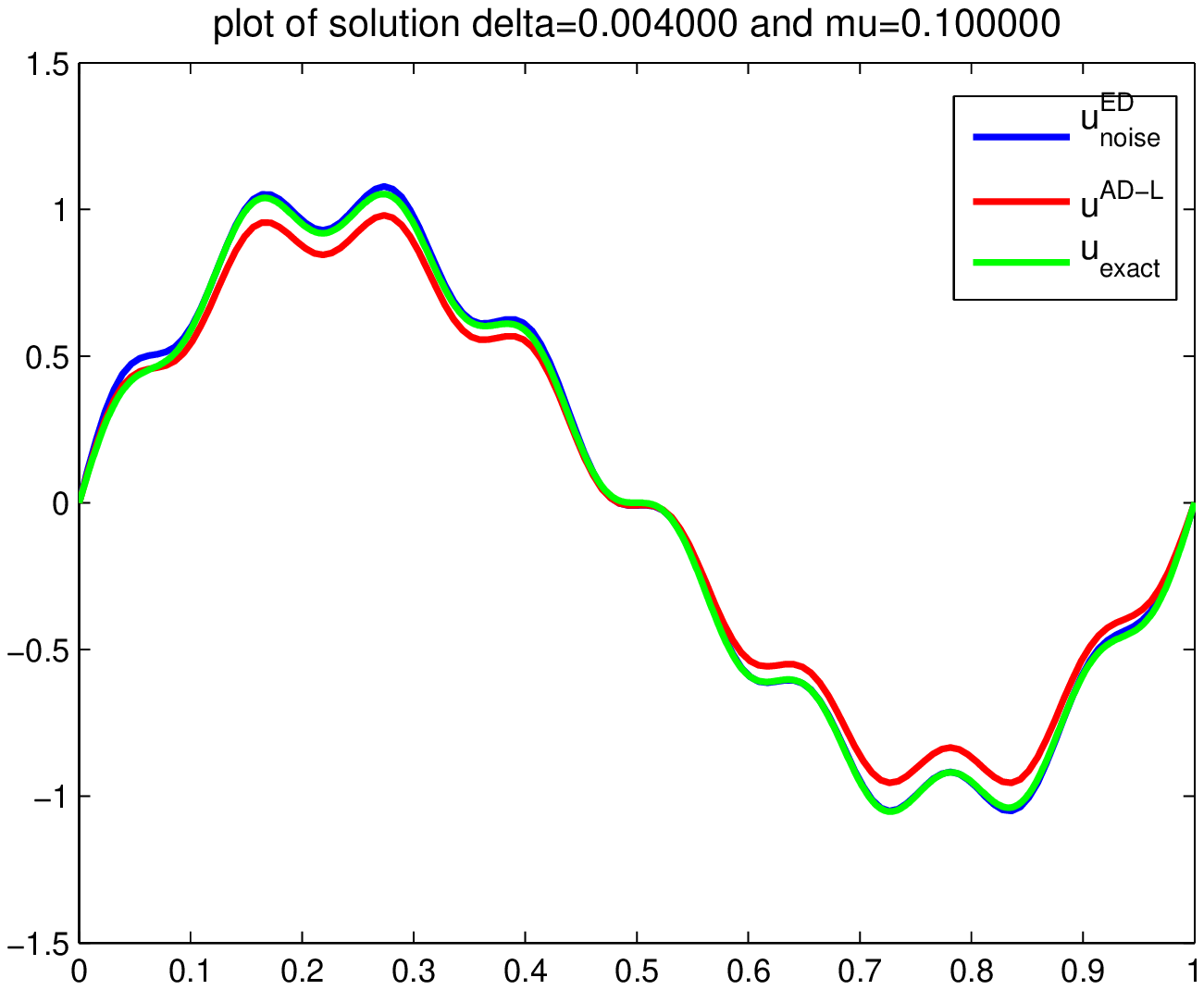}
\endminipage\hfill
\minipage{0.33\textwidth}
  \includegraphics[width=\linewidth]{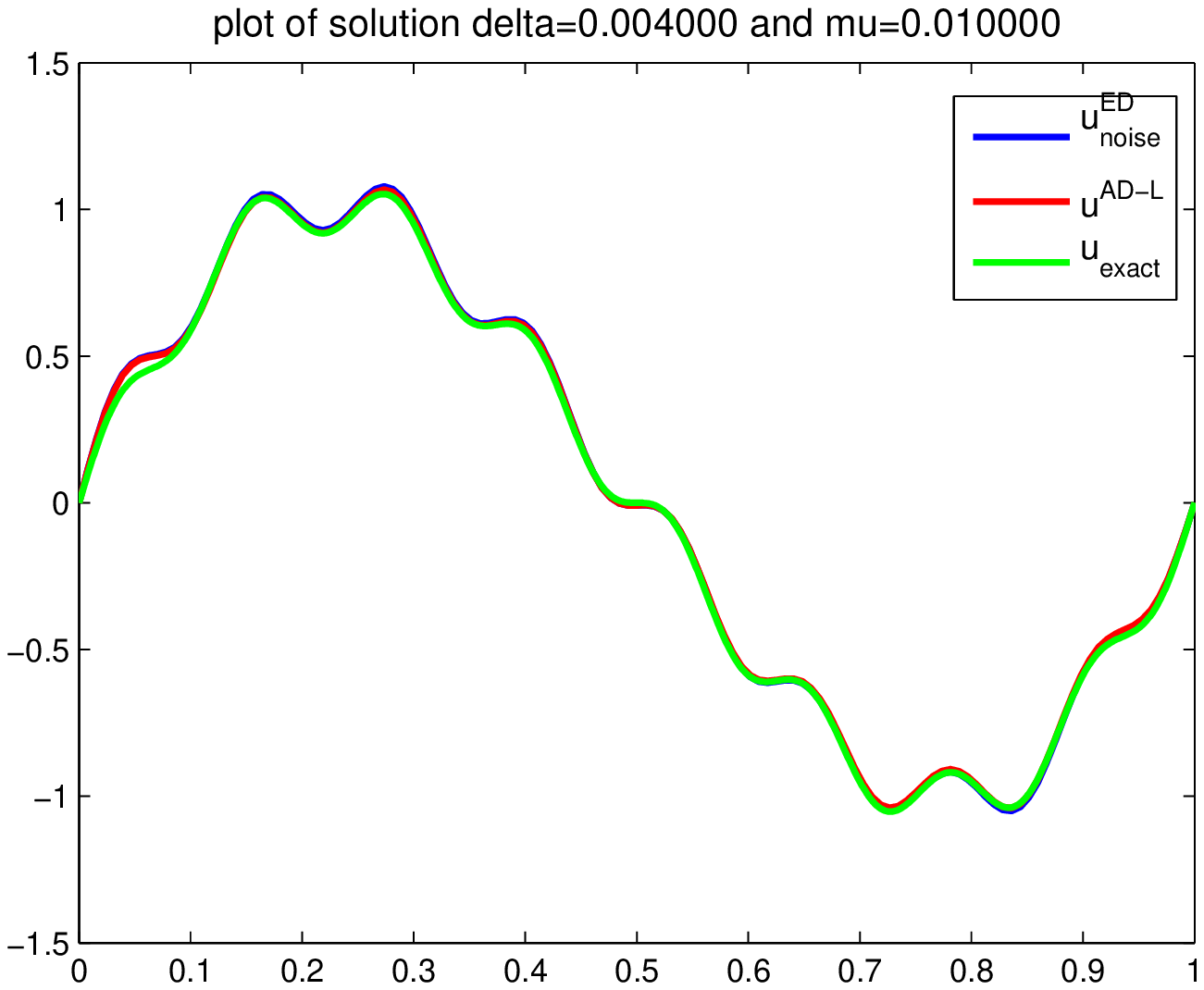}
\endminipage\hfill
\minipage{0.33\textwidth}
  \includegraphics[width=\linewidth]{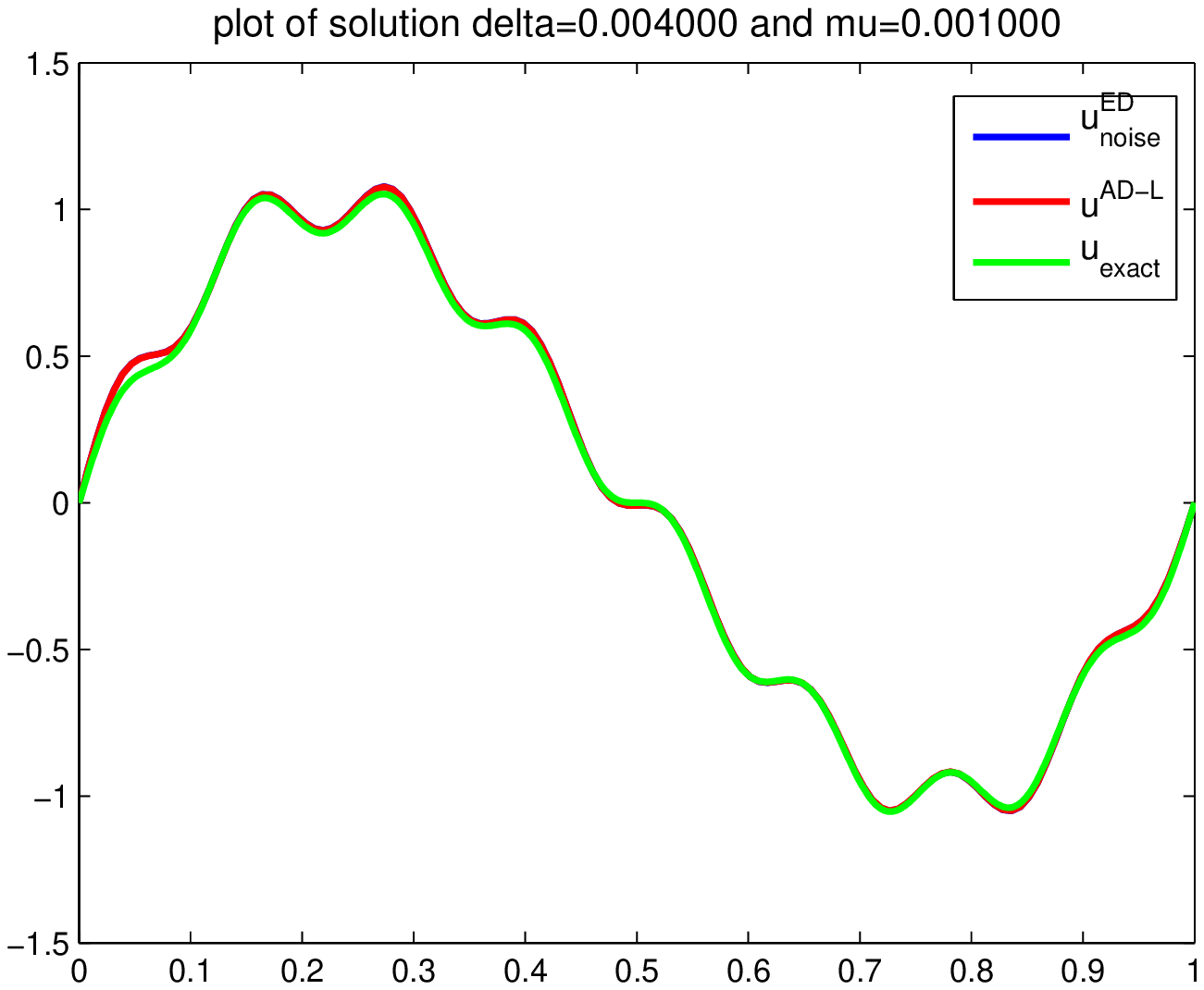}
\endminipage\hfill \\
\minipage{0.33\textwidth}
  \includegraphics[width=\linewidth]{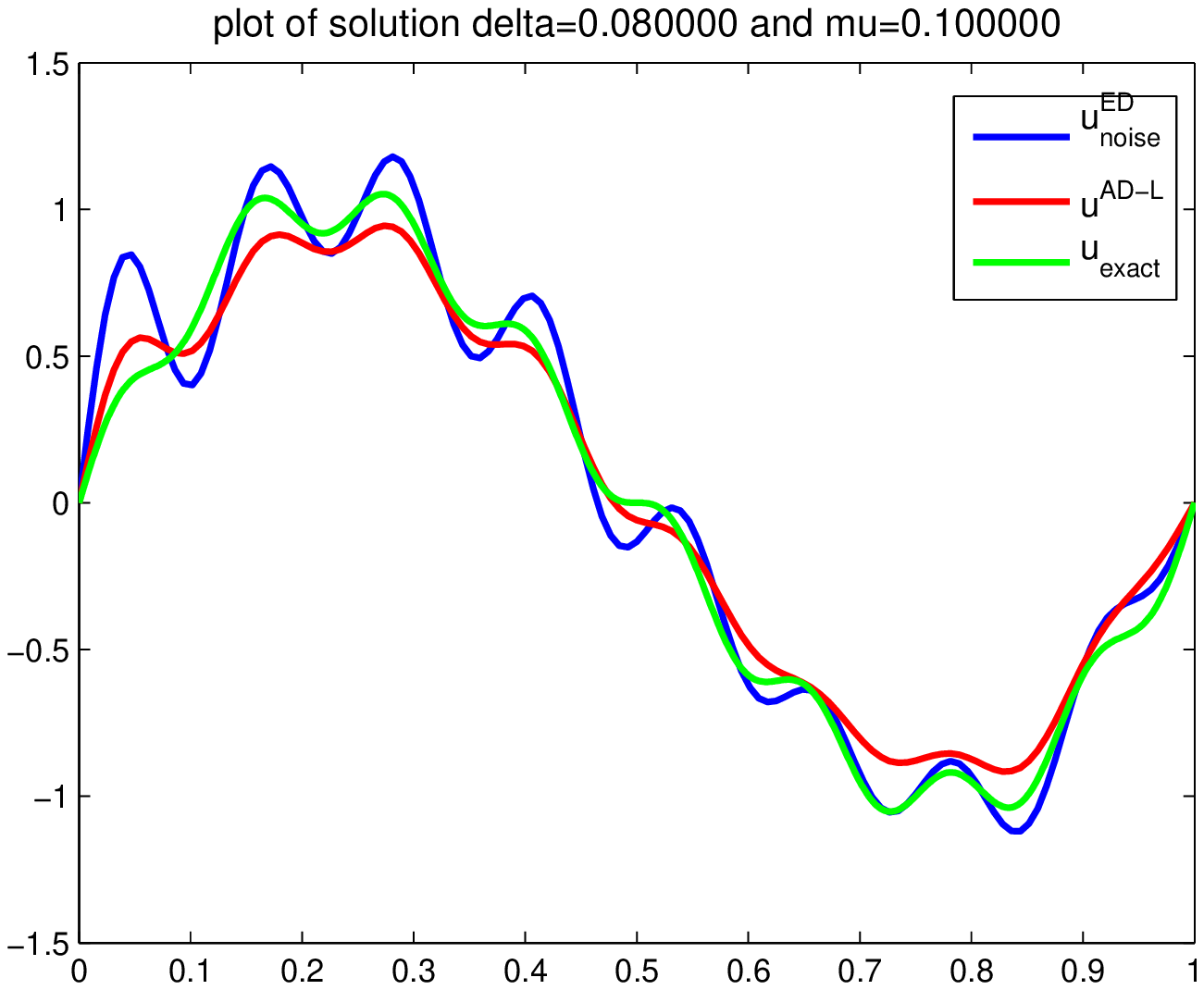}
\endminipage\hfill
\minipage{0.33\textwidth}
  \includegraphics[width=\linewidth]{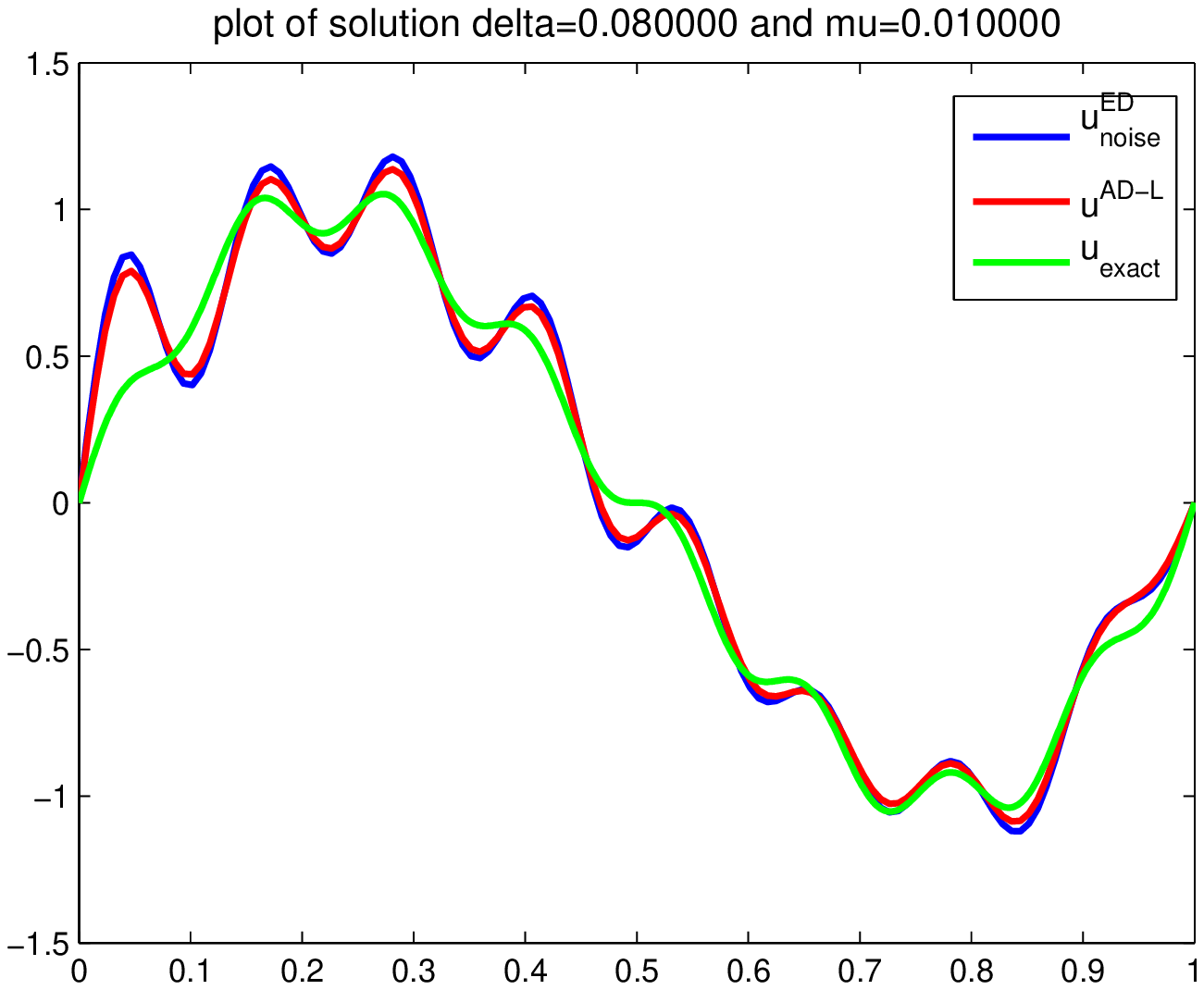}
\endminipage\hfill
\minipage{0.33\textwidth}
  \includegraphics[width=\linewidth]{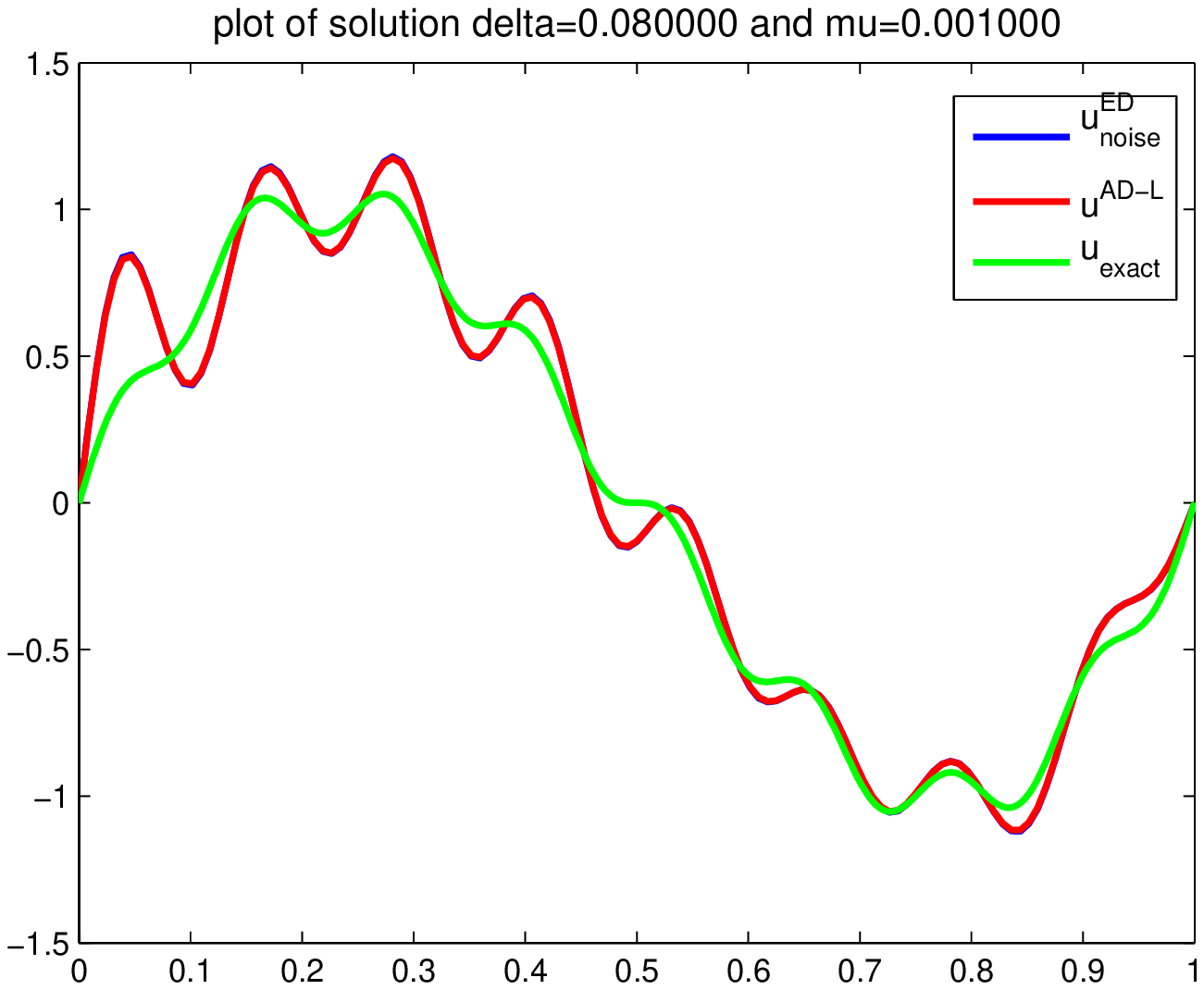}
\endminipage\hfill \\
\minipage{0.33\textwidth}
  \includegraphics[width=\linewidth]{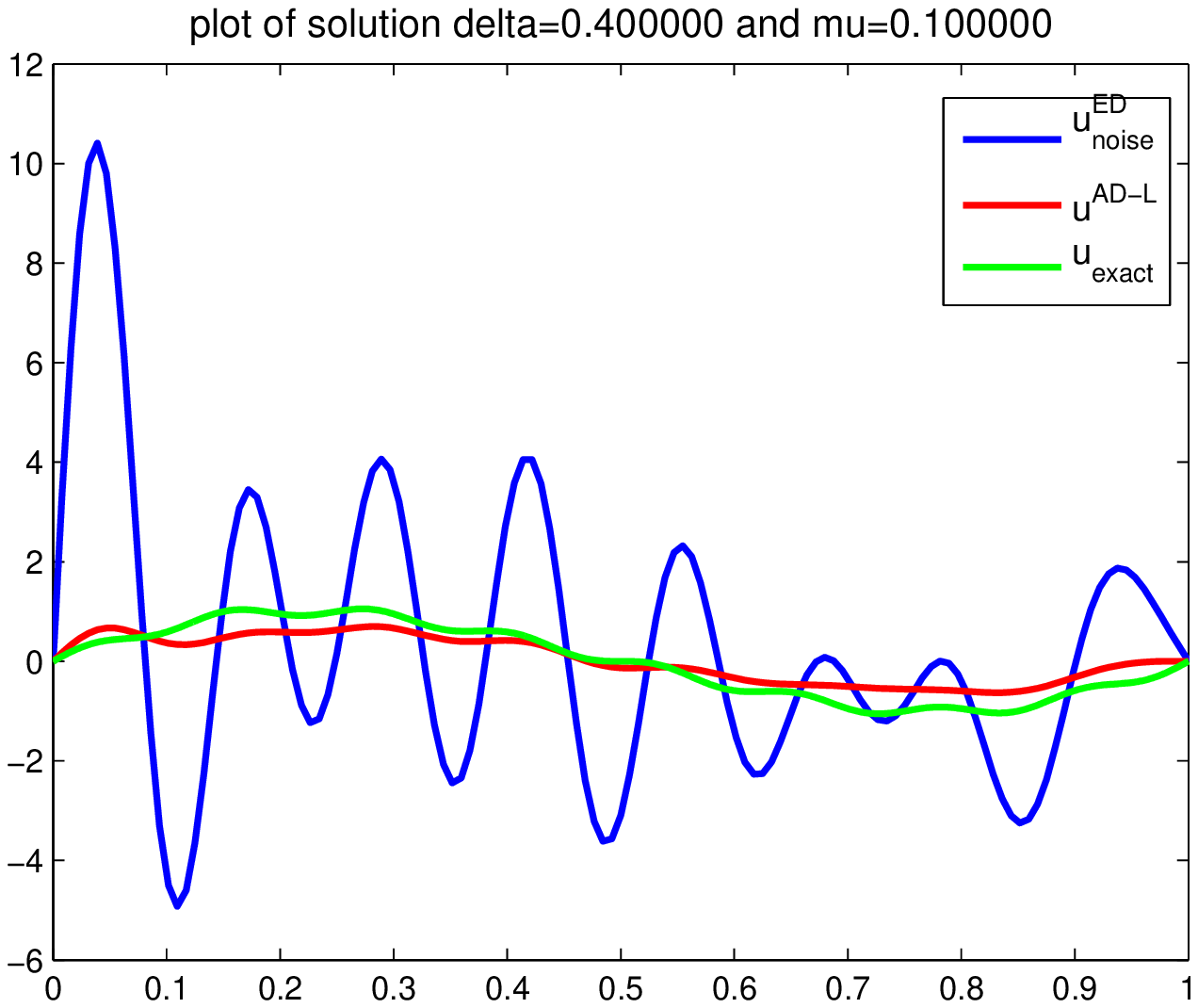}
\endminipage\hfill
\minipage{0.33\textwidth}
  \includegraphics[width=\linewidth]{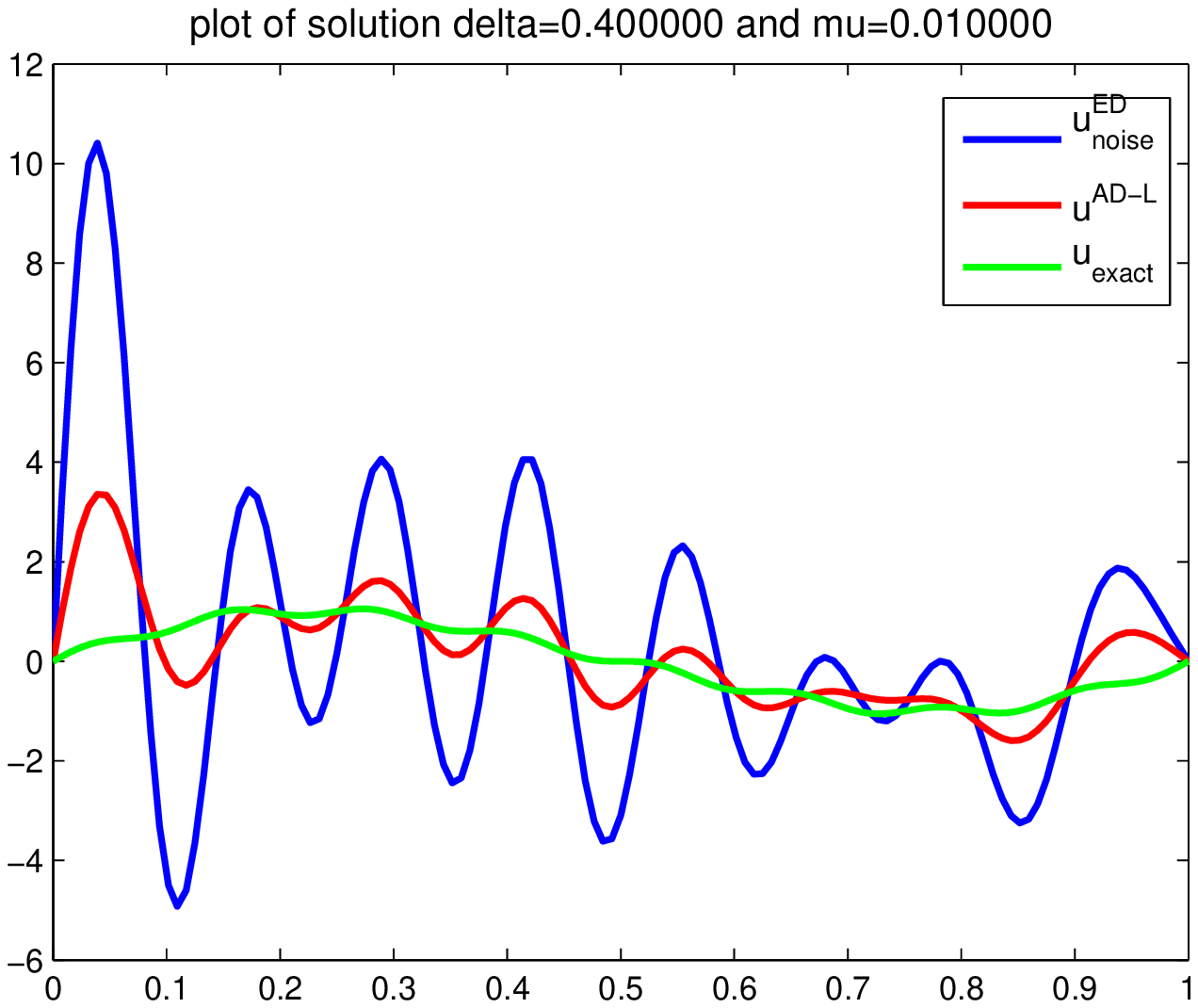}
\endminipage\hfill
\minipage{0.33\textwidth}
  \includegraphics[width=\linewidth]{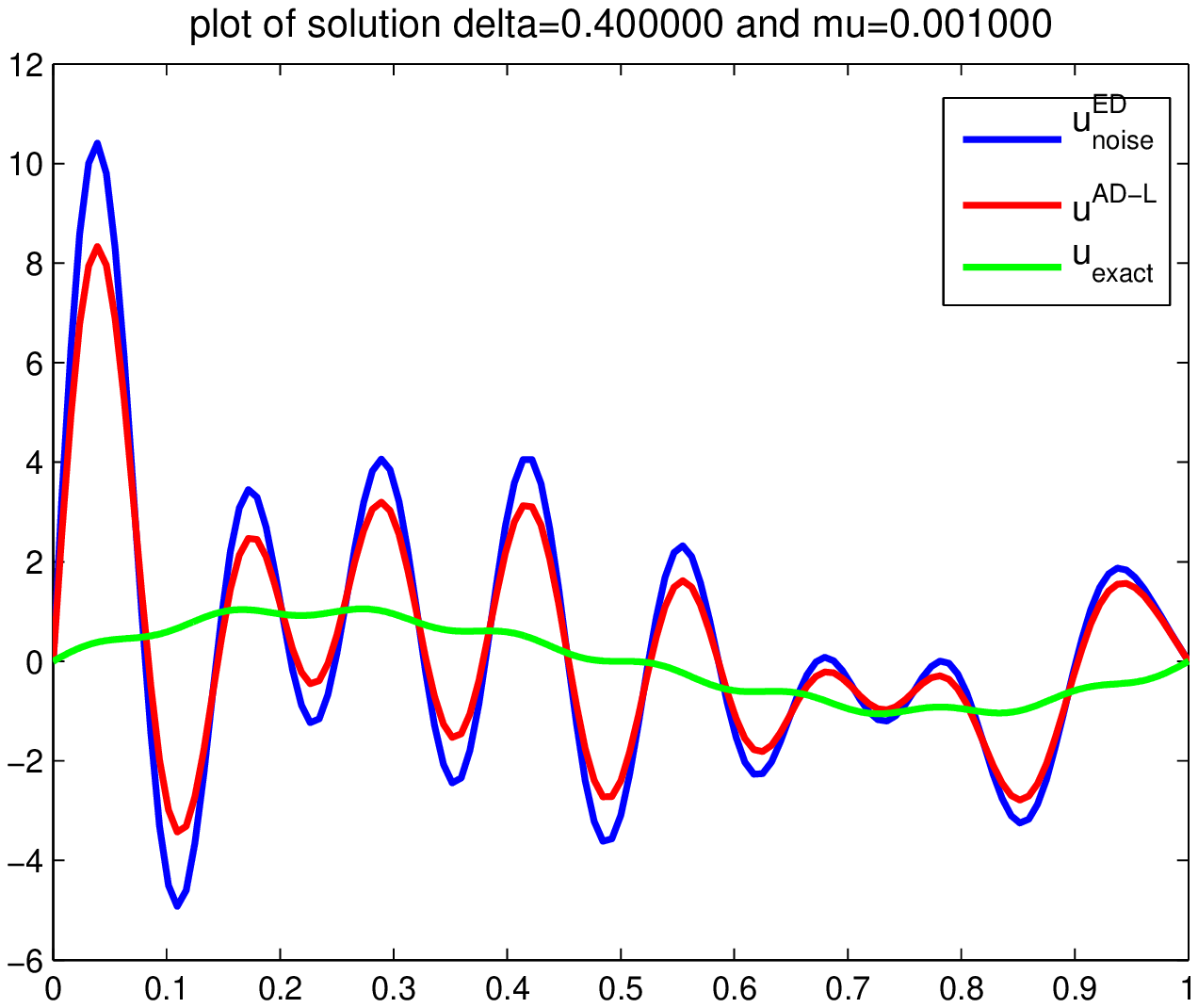}
\endminipage\hfill
	\caption{
		ROM exact deconvolution ($u^{ED}$, blue curve) and ROM approximate deconvolution with Lavrentiev regularization ($u^{AD-L}$, red curve) of a noisy filtered signal.
		A random noise with magnitude level $\mathcal{O}(10^{-2})$ is used.
		Three different $\delta$ values and three different $\mu$ values are used:
		$\delta = 0.004$ (top row), $\delta = 0.08$ (middle row), $\delta = 0.04$ (bottom row), $\mu = 0.1$ (left column), $\mu = 0.01$ (middle column) and $\mu = 0.001$ (right column).    
		The exact solution ($u$, green curve) is also plotted for comparison purposes.
		\label{fig:rom-ad-1}
	}
\end{figure}

\paragraph{Numerical Results}
As in Section~\ref{sec:rom-ed}, Algorithm 2 is used with the following parameter choices:
The input signal is  $u(x) = \sin(2 \, \pi \, x)  + 0.1 \sin(4 \, \pi \, x) + 0.1 \sin(16 \, \pi \, x)$ and the POD basis is $\{ \sin(\pi \, x), \sin(2 \, \pi \, x), \ldots \sin(16 \, \pi \, x) \}$, i.e., the Fourier basis.
A random noise with magnitude order $\mathcal{O}(10^{-2})$ is considered in Step 3 of Algorithm 2. 
(For clarity, results for the other two noise levels considered in Section~\ref{sec:rom-ed} are not presented in this section.)
The ROM exact deconvolution approximation ($u^{ED}$) and the ROM approximate deconvolution approximation with Lavrentiev regularization ($u^{AD-L}$) are plotted in Fig.~\ref{fig:rom-ad-1} for three representative $\delta$ values and three parameter $\mu$ values in the Lavrentiev regularization.
As benchmark, the exact input signal ($u$) is also plotted.
The plots in Fig.~\ref{fig:rom-ad-1} show that, as $\delta$ increases, $u^{AD-L}$ becomes significantly more accurate than $u^{ED}$.
Furthermore, larger $\mu$ values (i.e., larger regularization levels in the Lavrentiev regularization) increase the $u^{AD-L}$ accuracy.
These conclusions are reinforced by the errors of $u^{ED}$ and $u^{AD-L}$ listed in Table~\ref{table:rom-ad-1}.

To explain the increase in accuracy of $u^{AD-L}$ over $u^{ED}$, the ROM exact deconvolution condition number ($\mK^{ED}$ defined in~\eqref{eqn:rom-ad-ill-posed-1})and ROM approximate deconvolution with Lavrentiev regularization condition number $\mK^{AD-L} := \mK(\bM + \mu \, \bM + \mu \, \delta^2 \, \bS)$ are listed in Table~\ref{table:rom-ad-2}.
As $\delta$ increases, $\mK^{AD-L}$ gets smaller than $\mK^{ED}$.
Thus, the noise in the input filter signal is amplified more in the ROM exact deconvolution than in the ROM approximate deconvolution.

Given the scaling in~\eqref{eqn:rom-ad-ill-posed-2}, these numerical results are natural: an increase in $\delta$ results in an increase in $\mK^{ED}$, which in turn triggers an amplification of the input noise in the ROM exact deconvolution.
The same scaling in~\eqref{eqn:rom-ad-ill-posed-2} suggests that increasing $r$ will have an effect similar to that of increasing $\delta$.
To test this conjecture, we consider the signal $u(x) = \sin(\pi \, x)  + 0.1 \sin(50 \, \pi \, x) + 0.1 \sin(100 \, \pi \, x)$ and the POD basis is taken to be the Fourier basis $\{ \sin(\pi \, x), \sin(2 \, \pi \, x), \ldots \sin(100 \, \pi \, x) \}$.
Thus, we replaced $r=16$ with $r=100$ in our analytical example.
In this case, $\mK^{ED} = 350$, whereas $\mK^{AD-L} = 34.02$.
Furthermore, the $u^{ED}$ error is $11.0582$, which is more than $40$ times larger than the $u^{AD-L}$ error, which is $0.3113$.
These results support our conjecture that increasing $r$ will have an effect similar to that of increasing $\delta$.

Based on the results in Fig.~\ref{fig:rom-ad-1}, Table~\ref{table:rom-ad-1} and Table~\ref{table:rom-ad-2}, we conclude that AD-L is more accurate than ED.

\begin{table}[h]
\centering
\begin{tabular}{|l|c|c|c|c|l|}
\hline
\multicolumn{2}{|c|}{{ }} & $\mu = 0.1$ &  $\mu = 0.01$ & $\mu = 0.001$   \\ \hline
              &  $\|u-u^{ED}\|_{L^2}$            & $\|u-u^{AD-L}\|_{L^2}$ & $\|u-u^{AD-L}\|_{L^2}$ & $\|u-u^{AD-L}\|_{L^2}$  \\ \hline
   $\delta=0.004$ & 0.0167 & 0.0598 & 0.0148 & 0.0164  \\ \hline
   $\delta=0.08$    & 0.1214 & 0.0924 & 0.1035 & 0.1193  \\ \hline
   $\delta=0.4$   & 2.7389 & 0.2999 & 0.7749 & 2.1214  \\ \hline
\end{tabular}
	\caption{
		ROM exact deconvolution and ROM approximate deconvolution with Lavrentiev regularization.
		A random noise with magnitude level $\mathcal{O}(10^{-2})$ is used.
		Errors of $u^{ED}$ (second column) and $u^{AD-L}$ for three different $\mu$ values: $\mu = 0.1$ (third column), $\mu = 0.01$ (fourth column) and $\mu = 0.001$ (fifth column).
		Three different $\delta$ values are used.
	}
\label{table:rom-ad-1}
\end{table}

\begin{table}[h]
\centering
\begin{tabular}{|l|c|c|c|c|l|}
\hline
              &  $\mK^{ED}$           & $\mK^{AD-L}, \, \mu = 0.1$ & $\mK^{AD-L}, \, \mu = 0.01$ & $\mK^{AD-L}, \, \mu = 0.001$  \\ \hline
   $\delta=0.004$ & 1.01 & 1.02 & 1.02 & 1.02   \\ \hline
   $\delta=0.08$    & 15.93 & 2.41 & 1.13 & 1.00  \\ \hline
   $\delta=0.4$   & 155.13 & 32.58 & 4.85 & 1.31  \\ \hline
\end{tabular}
	\caption{
		Condition numbers for the ROM exact deconvolution ($\mK^{ED}$) and ROM approximate deconvolution with Lavrentiev regularization ($\mK^{AD-L}$).
		A random noise with magnitude level $\mathcal{O}(10^{-2})$ is used.
		Three $\delta$ values and three $\mu$ values are used.
			}
\label{table:rom-ad-2}
\end{table}

\clearpage

\section{Numerical Results}
\label{sec:numerical-results}

For a given input signal, we showed that the ROM exact deconvolution is an ill-posed problem that yields highly oscillatory results (Section~\ref{sec:rom-ed}), whereas the ROM approximate deconvolution with Lavrentiev regularization yields accurate results (Section~\ref{sec:rom-ad}).
In this section, we compare the ROM exact deconvolution and the ROM approximate deconvolution in the numerical discretization of the actual AD-ROM~\eqref{eqn:ad-rom}.
To this end, the deconvolution variable $\bw^{AD}_r$ in the AD-ROM is taken to be (i) the ROM exact deconvolution in Algorithm 1', and (ii) the ROM approximate deconvolution in Algorithm 2. 
By convention, in the remainder of the paper, the AD-ROM~\eqref{eqn:ad-rom} is called ED-ROM in case (i) and AD-ROM in case (ii).
Specifically, for the forward Euler time discretization used in this paper, the following algorithm needs to be used at each time step for ED-ROM and AD-ROM:
Given $\bw^{n}_r$, which is the approximation of $\obu_r$ at the current time step $n$, find $\bw^{n+1}_r$, which is the approximation of $\obu_r$ at the next time step $n+1$ as follows:

\begin{center}
	{\bf Algorithm 3}
\end{center}
\begin{itemize}
	\item[(1)]
		\begin{equation}
		    \hspace*{-1.0cm}
			\bw^{D,n}_r =
			\begin{cases}
   				\bw_r^{ED,n} = (\bM + \delta^2 \, \bS) \, (\bw_r^{n} + \bfeta) & \text{for ED-ROM} \\
   				\bw_r^{AD-L,n} = (\bM + \mu \, \bM + \mu \, \delta^2 \, \bS)^{-1}(\bM + \delta^2 \, \bS) \, (\bw_r^{n} + \bfeta) & \text{for AD-ROM} \, .
  			\end{cases}
		\label{eqn:numerical-results-1}
	\end{equation}
\item[(2)]
	    \begin{equation}
       \left(
            \frac{\bw_r^{n+1} - \bw_r^{n}}{\Delta t} , \bphi_{k}
        \right)
        + Re^{-1} \, \left(
            \nabla \bw_r ,
            \nabla \bphi_{k}
        \right)
        + \biggl(
            {\overline{\bw^{D,n}_r \, \bw^{D,n}_r}} ,
            \nabla \bphi_{k}
        \biggr) 
        = 0 \, ,
    \label{eqn:ad-rom-discretization}
    \end{equation}
	where $\bfeta$ is the noise and $\Delta t$ is the time step.
\end{itemize}

For both the ED-ROM and AD-ROM, we need to filter the initial condition: $\bw_r(\bx,0) = \overline{\bu_0}(\bx)$.
We also need to filter the boundary conditions.
Since we are using the DF~\eqref{eqn:df}, the boundary conditions remain homogeneous Dirichlet: $\bw_r(\bx,t) = {\bf 0}$ on the boundary. 

To compare the AD-ROM with the ED-ROM, as in Sections~\ref{sec:rom-ed} and \ref{sec:rom-ad}, noise of two different magnitudes ($\mO(10^{-2})$ and $\mO(10^{-3})$) is used in Algorithm 3.

\begin{remark}[Noise Modeling AD-ROM Discretization Error]
We emphasize that the addition of noise to Algorithm 3 is relevant to ROM of {\it realistic} flows, e.g., structure dominated turbulent flows~\cite{HLB96}, which represent the ultimate target of the new AD-ROM.
In these realistic settings, noise could model inaccuracies in the data (e.g., forcing term or boundary conditions) or numerical inaccuracies.
In this paper, we exclusively consider that noise models numerical inaccuracies (although the other scenario is equally important).
We note, however, that the numerical investigation in Section~\ref{sec:numerical-results} is carried out  for the 1D Burgers equation. 
In this simple setting, we can afford to use a fine spatial and temporal resolution and a relatively high number of POD modes. 
Thus, the numerical error is small in this case.
However, to get some insight into the new AD-ROM's performance in realistic flows, in which the numerical error is large, we add noise to Algorithm 3. 
We emphasize, however, that we ensure that the magnitude of the noise added to Algorithm 3 is of the same (or lower) order as the magnitude of the numerical discretization error in the AD-ROM.

Indeed, the total discretization error in the AD-ROM has three main components~\cite{iliescu2014are}: the time discretization error (which, for the forward Euler method used in this section, is $\mO(\Delta t) = \mO(10^{-4})$), the spatial discretization error (which, for the linear FE method used in this section, is $\mO(h^2) = \mO(10^{-6})$), and the POD truncation error (which, for the $r$ values used in this section, is $\mO(\sqrt{\sum_{i=r+1}^{d} \lambda_i}) = \mO(10^{-2})$ or lower).
Thus, the magnitude of AD-ROM's total discretization error is $\mO(10^{-2})$, which is of the same order as (or higher than) the magnitude of the noise used in Algorithm 3.
\label{remark:noise}
\end{remark}

In the remainder of the section, we compare the AD-ROM with the ED-ROM.
For comparison purposes, results for the G-ROM and DNS are also included.
Numerical results are presented for the Burgers equation (Section~\ref{sec:numerical-results-burgers}).

\subsection{Burgers Equation}
\label{sec:numerical-results-burgers}

        The mathematical model used in this section is the (1D)
        Burgers equation
        \begin{eqnarray}
          \left\{\begin{array}{ll}
          u_t-\nu \, u_{xx}+u \, u_x = 0& x\in [0,1],                         \\
          u(x,0)=u_0(x)& x\in [0,1],                                          \\
          u(0,t)=u(1,t) = 0 \, , &
          \end{array} \right.
        \label{eqn:burgers}
         \end{eqnarray}
        where $\nu=10^{-3}$ is the diffusion parameter and $t\in[0,1]$. The
        initial condition is $u_0(x)=1$ for $x\in(0,\frac{1}{2}]$ and $u_0(x)=0$
        for $x\in(\frac{1}{2},1)$. This computational setting is similar to that
        used in~\cite{KV01,wells2015regularized}.

        For the DNS, a uniform mesh with $h=1/1024$ and a linear FE were
        used for the spatial discretization and the forward Euler method with a
        time step $\Delta t=10^{-4}$ were used for the time discretization. 
        For ROMs, the forward Euler method with a time step
        $\Delta t=10^{-4}$ were used for the time discretization. A total of
        $101$ snapshots were collected and the following $r$ values were used: $r = 5, 10, 15$ and $20$.

        For all the parameters used in the numerical investigation, the CPU
        times of the ED-ROM and AD-ROM were significantly lower that
        the CPU time of the corresponding DNS. The DNS took about \(200\)
        seconds to run. 
In contrast, the CPU times of the G-ROM and the ED-ROM and AD-ROM were on the order of $10$ seconds.

The ED-ROM and AD-ROM errors are listed in Table~\ref{tab:ad-rom-1} for four $r$ values, two $\delta$ values and two noise magnitudes.
The optimal $\mu$ values were used in the AD-ROM. 
For comparison purposes, G-ROM errors are also listed.
To compute the errors, the ED-ROM and AD-ROM results are compared with the filtered DNS data, whereas the G-ROM results are compared with the unfiltered DNS data.
The results in Table~\ref{tab:ad-rom-1} show that the AD-ROM is consistently more accurate than the ED-ROM.
In particular, for large $r$ values, large $\delta$ values and large noise levels, the AD-ROM error is as much as five times lower than the ED-ROM error.
This is due to the fact that $\mK^{AD-L}$ is significantly lower than $\mK^{ED}$, especially for high $\delta$ and $r$ values (see Table~\ref{tab:ad-rom-2}). 

We note that the G-ROM is less accurate than the ED-ROM and AD-ROM for $r=5$, but more accurate for the larger $r$ values.
We emphasize, however, that the G-ROM yields accurate results for the relatively simple Burgers equation considered in this section.
For realistic flows, significantly higher $r$ values are generally needed~\cite{osth2014need,protas2015optimal} and, thus, we expect the ED-ROM and AD-ROM to perform significantly better than G-ROM~\cite{wang2012proper}.

\begin{figure}[h]
\includegraphics[width=3in]{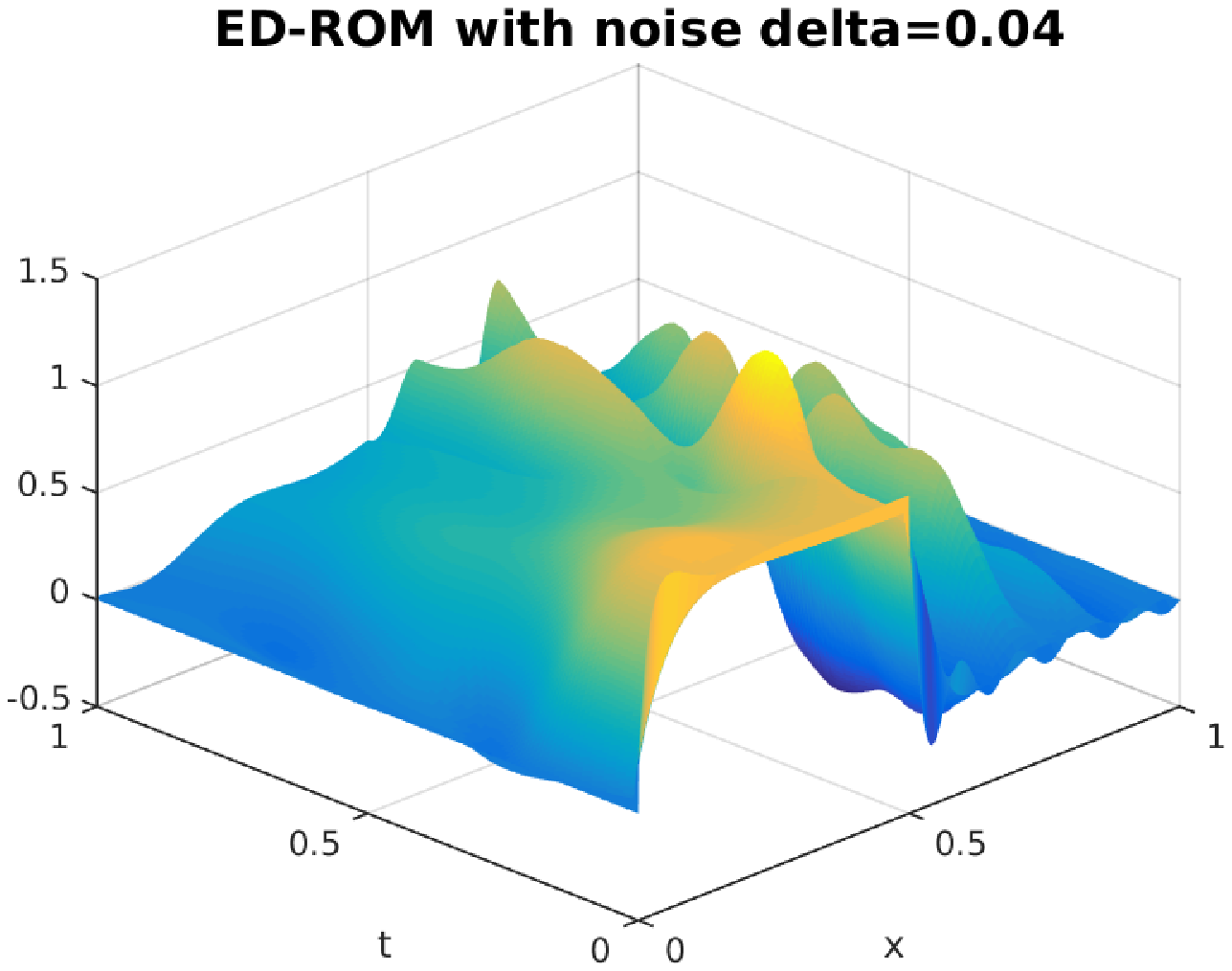}
\includegraphics[width=3in]{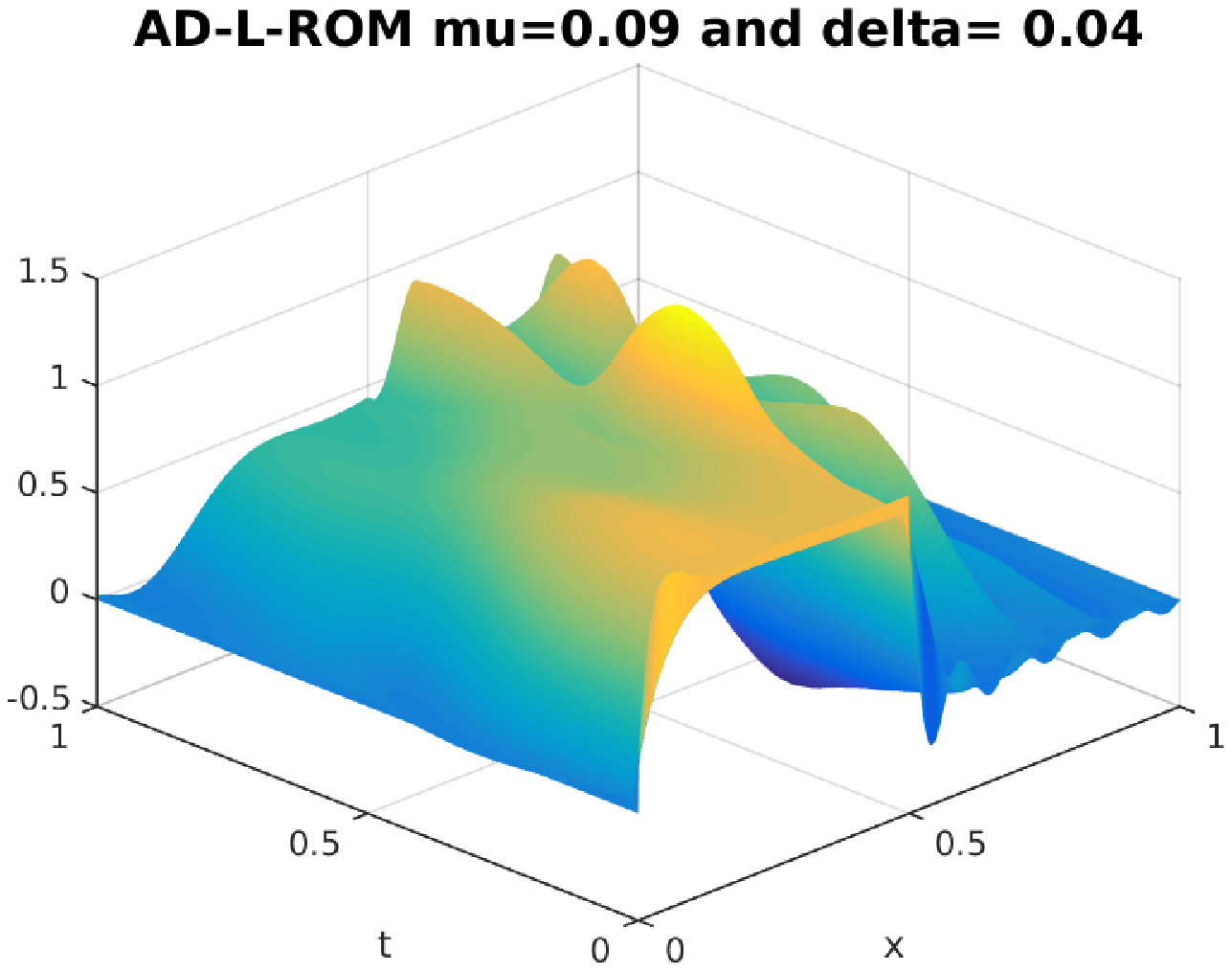}\\
\includegraphics[width=3in]{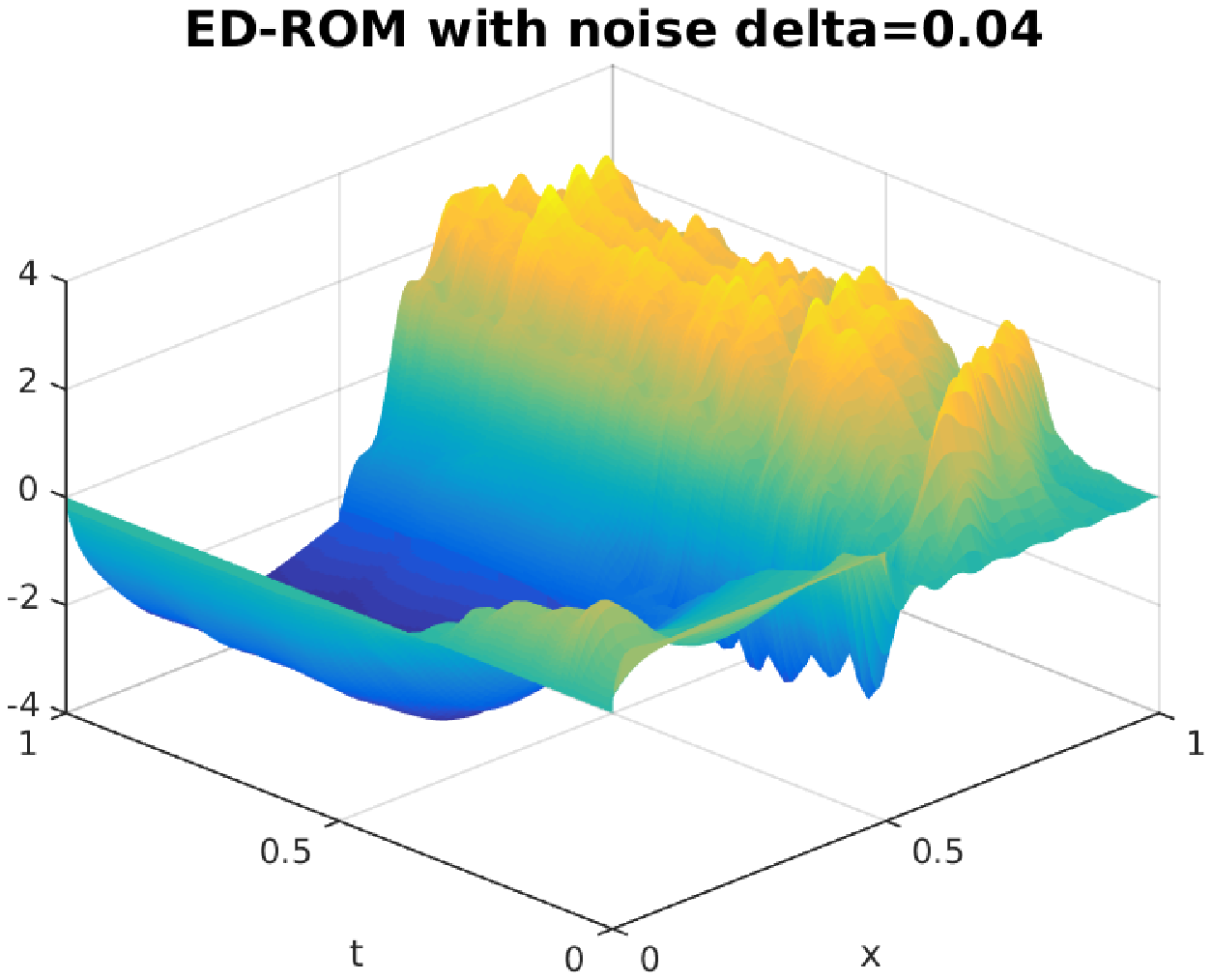}
\includegraphics[width=3in]{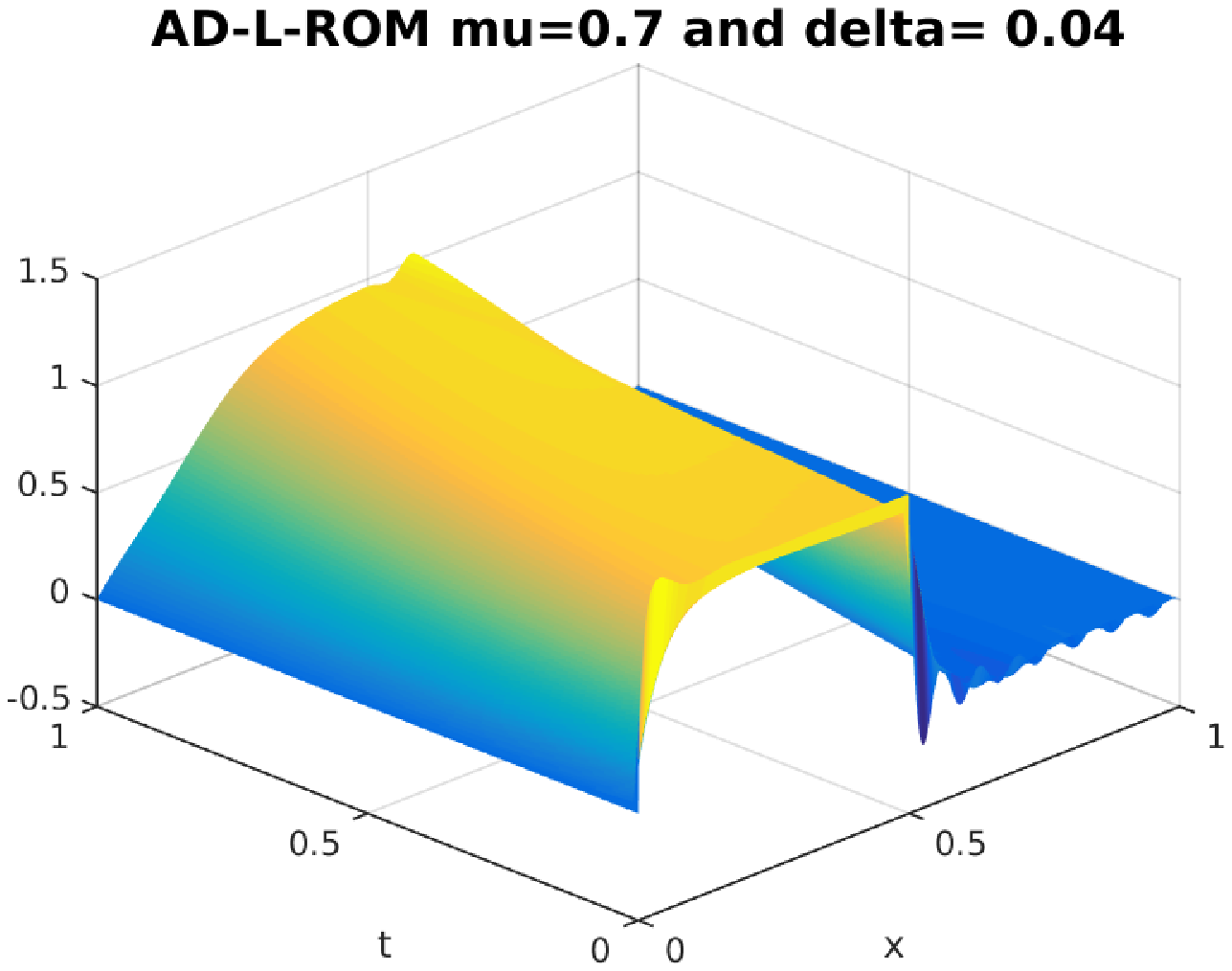}\\
\includegraphics[width=3in]{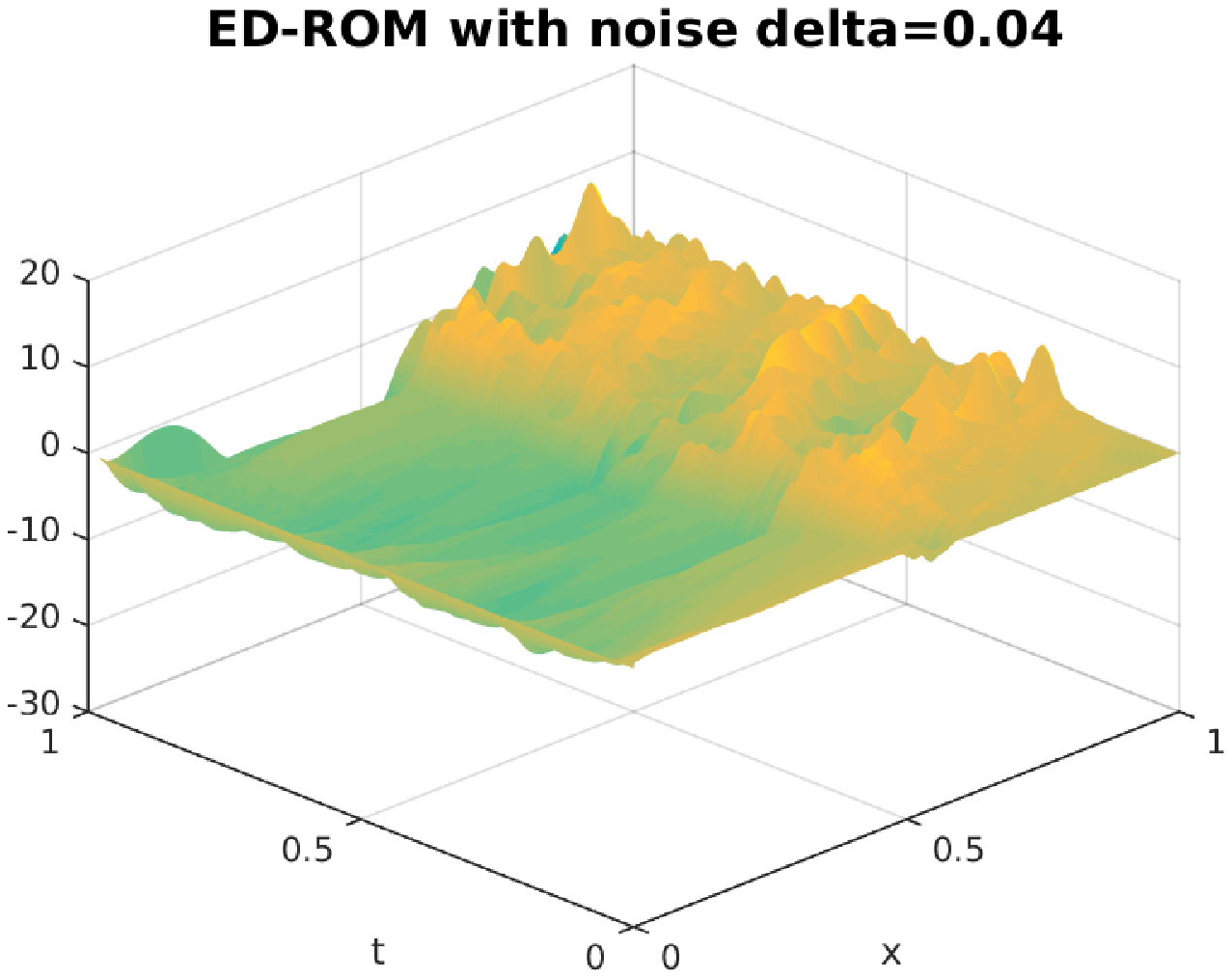}
\includegraphics[width=3in]{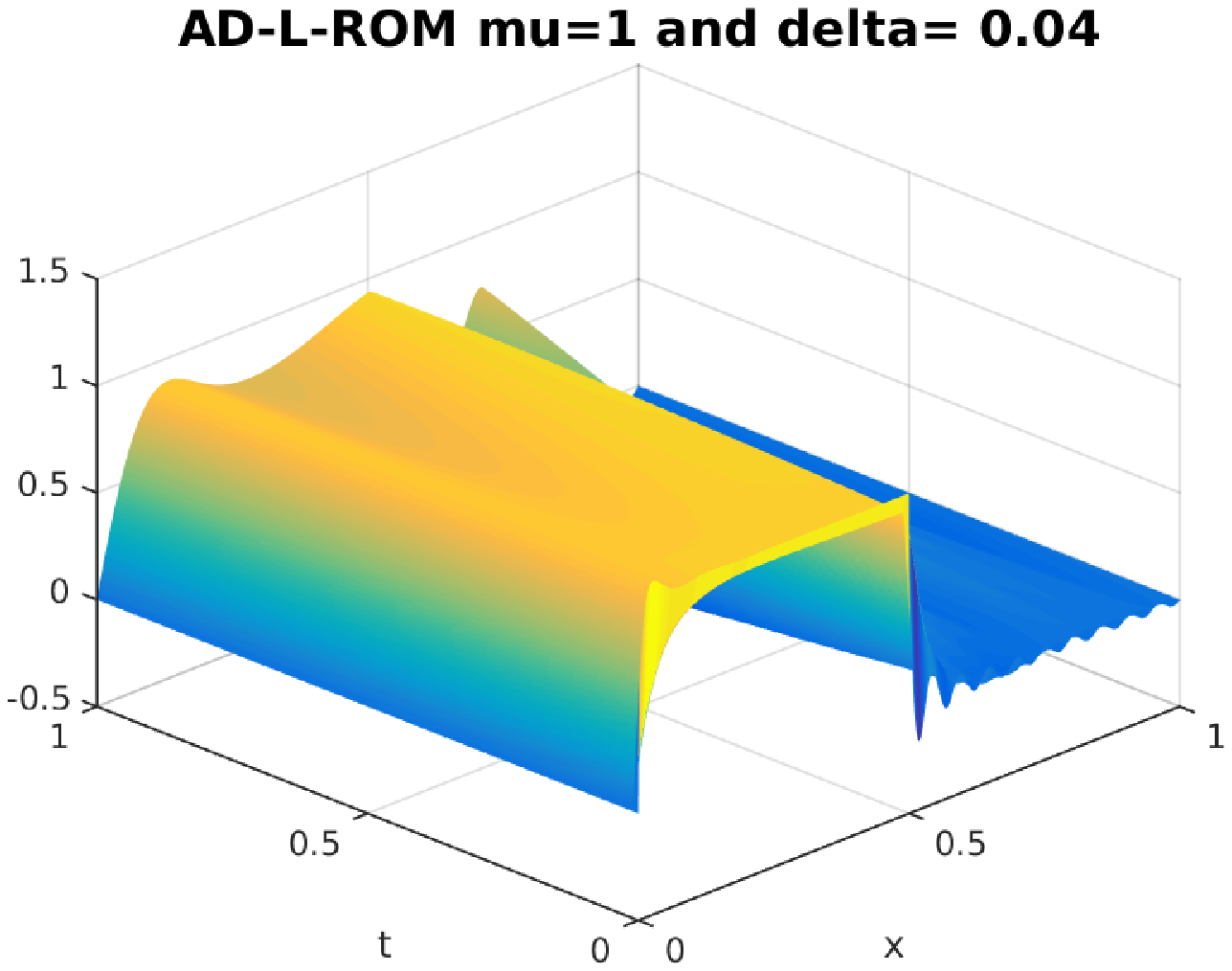}\\
	\caption{
		Burgers equation, ED-ROM (left column) and AD-ROM (right column) plots for $\delta=0.04$, noise magnitude $10^{-2}$, and $r=10$ (top row), $r=15$ (middle row) and $r=20$ (bottom row).
		\label{fig:ad-rom-1}
	}
\end{figure}

The results in Table~\ref{tab:ad-rom-1} are supported by the plots in Fig.~\ref{fig:ad-rom-1}.

\begin{table}[htp]
\centering
\begin{tabular}{|c|c|c|c|c|c|c|c|}
\hline
\multirow{2}{*}{r} & \multirow{2}{*}{$\delta$} & \multicolumn{2}{c|}{G-ROM} & \multicolumn{2}{c|}{ED-ROM}  &  \multicolumn{2}{c|}{AD-ROM}   \\ \cline{3-8}
           &                   &  $\mO(10^{-2})$  & $\mO(10^{-3})$   & $\mO(10^{-2})$  & $\mO(10^{-3})$  & $\mO(10^{-2})$  & $\mO(10^{-3})$      \\ \hline
\multirow{2}{*}{$r=5$} & $\delta=0.04$ & 0.2248 & 0.2007 & 0.1526 & 0.1233 & 0.1474 & 0.1167         \\ \cline{2-8} 
                  &  $\delta=0.004$& 0.2248 & 0.2007 & 0.2219 & 0.1964 & 0.1871& 0.1841        \\ \hline
\multirow{2}{*}{$r=10$} &       $\delta=0.04$  & 0.2784  & 0.1679 & 0.3435 & 0.0771 & 0.2759 & 0.0718        \\ \cline{2-8} 
                  &  $\delta=0.004$ &   0.2784 & 0.1679 & 0.2789 & 0.1604 & 0.2105 & 0.1487         \\ \hline
\multirow{2}{*}{$r=15$} & $\delta=0.04$  & 0.2245 & 0.0905 & 2.0121 & 0.1513 & 0.4208 & 0.1483                 \\ \cline{2-8} 
                  & $\delta=0.004$  &  0.2245   & 0.0905 & 0.2214 & 0.0800 & 0.2117 & 0.0721      \\ \hline
\multirow{2}{*}{$r=20$} &   $\delta=0.04$  & 0.2589  & 0.0610 & NA & 0.2813 & 0.4108 & 0.1284      \\ \cline{2-8} 
                  &   $\delta=0.004$ & 0.2589 & 0.0610 & 0.2634 & 0.0562 & 0.2361 & 0.0472  \\ \hline
\end{tabular}
	\caption{
		Burgers equation, ED-ROM and AD-ROM errors for four $r$ values, two $\delta$ values and two noise magnitude levels.
		G-ROM errors are also listed for comparison purposes.
		\label{tab:ad-rom-1}
	}
\end{table}

\begin{table}[htp]
\centering
\begin{tabular}{|c|c|c|c|}
\hline
  r      & $\delta$ & $\mK^{ED}$ & $\mK^{AD-L}$ \\ \hline
\multirow{2}{*}{$r=5$} & $\delta=0.04$  & 2.80  & 1.01 \\ \cline{2-4} 
                  & $\delta=0.004$ & 1.01 & 1.00 \\ \hline
\multirow{2}{*}{$r=10$} & $\delta=0.04$ & 14.10 & 1.74 \\ \cline{2-4}
                  & $\delta=0.004$ & 1.13 & 1.01 \\ \hline
\multirow{2}{*}{$r=15$} & $\delta=0.04$ & 43.61 & 12.55\\ \cline{2-4}
                  & $\delta=0.004$ &  1.50 & 1.01 \\ \hline
\multirow{2}{*}{$r=20$} & $\delta=0.04$ & 101.59 & 33.83 \\ \cline{2-4}
                  & $\delta=0.004$ & 2.15 & 1.04 \\ \hline
\end{tabular}
	\caption{
		Burgers equation, ED-ROM and AD-ROM condition numbers for four $r$ values and two $\delta$ values.
		\label{tab:ad-rom-2}
	}
\end{table}

\clearpage

\clearpage

\section{Conclusions and Future Work}
\label{sec:conclusions}

Explicit ROM spatial filtering was used to develop a large eddy simulation ROM (LES-ROM) framework.
Within this LES-ROM framework, an approximate deconvolution ROM (AD-ROM) was proposed.
The AD-ROM was assessed in the numerical simulation of the Burgers equation with a small diffusion coefficient ($\nu = 10^{-3}$). 
Noise of different magnitude levels was used to model the numerical error.
Different values for the ROM spatial filter radius $\delta$ were used.
The $L^2$-norm of the error was the criterion used in the numerical assessment of the AD-ROM. 
The numerical investigation showed that the AD-ROM was significantly more accurate than the exact deconvolution ROM (ED-ROM), in which an exact filter inversion was employed.
Furthermore, the AD-ROM was more accurate than the standard G-ROM for low $r$ values, which were appropriate for the test problems employed in this paper.
It should be emphasized that the CPU time of the AD-ROM was orders of magnitude lower than the CPU time of the DNS.

These first steps in the investigation of the AD-ROM yielded encouraging results. 
There are, however, several research directions that could be further pursued. 
For example, one could consider different energy functionals~\cite{bertero1998introduction} in the AD regularization method employed. 
One could also investigate the magnitude of the commutation error in the AD-ROM (see Remark~\ref{remark:commutation-error}).
If the commutation error is relatively large compared with the discretization and modeling errors, then various modeling strategies could be sought, just as in LES~\cite{BIL05}.

Probably the important research direction that we plan to pursue is to investigate the AD-ROM in the numerical simulation of realistic structure dominated turbulent flows, such as those mentioned in the introduction (see, e.g., \cite{osth2014unsteady,osth2014need}).
In these realistic flows, the AD-ROM's numerical discretization error will play a more prominent role due to, e.g., the inherent coarse spatial mesh size, coarse time step and low number of POD modes used.
Thus, although for the simplified settings considered in this paper noise was artificially  added to the AD-ROM, in the realistic setting above noise will be naturally present in the form of numerical inaccuracies.
We note that, in~\cite{giere2015supg}, noise has already been shown to have a significant effect on the ROM accuracy for a convection-dominated convention-diffusion-reaction problem on an inherent coarse spatial mesh.
In realistic structure dominated turbulent flows, noise will probably have an even stronger effect on ROM accuracy.
In these settings, when numerical noise is naturally present, we expect that the AD-ROM (possibly with additional stabilization mechanisms) to clearly outperform the ED-ROM and the G-ROM.

\bibliographystyle{plain}
\bibliography{traian}

\end{document}